\documentclass[twocolumn]{aastex62}

\newcommand{\ts}{\thinspace}

\usepackage{rotating}
\usepackage{verbatimbox}

\graphicspath{{./}{figures/}}

\accepted{: August 2023}

\submitjournal{ApJ}

\shortauthors{Auge et al.}

\begin{document}

\title{The Accretion History of AGN: The Spectral Energy Distributions of X-ray Luminous AGN}

\correspondingauthor{Connor Auge}
\email{cauge.@hawaii.edu}

\author[0000-0002-5504-8752]{Connor Auge}
\affil{Institute for Astronomy, University of Hawai\`{}i, 2680 Woodlawn Drive, Honolulu, HI 96822, USA}

\author[0000-0002-1233-9998]{David Sanders}
\affiliation{Institute for Astronomy, University of Hawai\`{}i, 2680 Woodlawn Drive, Honolulu, HI 96822, USA}

\author[0000-0001-7568-6412]{Ezequiel Treister}
\affiliation{Instituto de Astrof\'isica, Facultad de F\'isica, Pontificia Universidad Cat\'olica de Chile, Casilla 306, Santiago 22, Chile}

\author[0000-0002-0745-9792]{C. Megan Urry}
\affiliation{Physics Department and Yale Center for Astronomy \& Astrophysics, PO Box 208120, New Haven, CT 06520-8120, USA}

\author[0000-0002-1306-1545]{Allison Kirkpatrick}
\affiliation{Department of Physics \& Astronomy, University of Kansas, Lawrence, KS 66045, USA}

\author[0000-0002-1697-186X]{Nico Cappelluti}
\affiliation{Department of Physics, University of Miami, Coral Gables, FL 33124, USA}

\author[0000-0001-8211-3807]{Tonima Tasnim Ananna}
\affiliation{Department of Physics and Astronomy, Dartmouth College, 6127 Wilder Laboratory, Hanover, NH 03755, USA}

\author[0000-0003-0946-6176]{Médéric Boquien}
\affiliation{Instituto de Alta Investigación, Universidad de Tarapacá, Casilla 7D, Arica, Chile}

\author[0000-0003-0476-6647]{Mislav Balokovi\'{c}}
\affiliation{Physics Department and Yale Center for Astronomy \& Astrophysics, PO Box 208120, New Haven, CT 06520-8120, USA}

\author[0000-0002-2115-1137]{Francesca Civano}
\affiliation{NASA Goddard Space Flight Center, Greenbelt, MD 20771, USA}

\author[0000-0003-3930-7950]{Brandon Coleman}
\affiliation{Department of Physics \& Astronomy, University of Kansas, Lawrence, KS 66045, USA}

\author[0000-0002-2525-9647]{Aritra Ghosh}
\affiliation{Department of Astronomy, Yale University, New Haven, CT, USA}
\affiliation{Yale Center for Astronomy and Astrophysics, New Haven, CT, USA}

\author[0000-0001-9187-3605]{Jeyhan Kartaltepe}
\affiliation{Laboratory for Multiwavelength Astrophysics, School of Physics
and Astronomy, Rochester Institute of Technology, 84 Lomb Memo-
rial Drive, Rochester, NY 14623, USA}

\author[0000-0002-7998-9581]{Michael Koss}
\affiliation{Eureka Scientific, Inc., 2452 Delmer Street, Suite 100, Oakland, CA 94602-3017, USA}
\affiliation{Space Science Institute, 4750 Walnut Street, Suite 205, Boulder, CO 80301, USA}

\author[0000-0002-5907-3330]{Stephanie LaMassa}
\affiliation{Space Telescope Science Institute, 3700 San Martin Drive, Baltimore, MD 21210, USA}

\author[0000-0001-5544-0749]{Stefano Marchesi}
\affiliation{Dipartimento di Fisica e Astronomia (DIFA), Università di Bologna, via Gobetti 93/2, I-40129 Bologna, Italy}
\affiliation{Department of Physics and Astronomy, Clemson University,  Kinard Lab of Physics, Clemson, SC 29634, USA}
\affiliation{INAF - Osservatorio di Astrofisica e Scienza dello Spazio di Bologna, Via Piero Gobetti, 93/3, 40129, Bologna, Italy}

\author[0000-0003-2196-3298]{Alessandro Peca}
\affiliation{Department of Physics, University of Miami, Coral Gables, FL 33124, USA}

\author[0000-0003-2284-8603]{Meredith Powell}
\affiliation{Kavli Institute for Particle Astrophysics and Cosmology, Stanford University, 452 Lomita Mall, Stanford, CA 94305}

\author[0000-0002-3683-7297]{Benny Trakhtenbrot}
\affiliation{School of Physics and Astronomy, Tel Aviv University, Tel Aviv 69978, Israel}

\author[0000-0003-2971-1722]{Tracey Jane Turner}
\affiliation{Eureka Scientific, Inc., 2452 Delmer Street, Suite 100, Oakland, CA 94602-3017, USA}

\begin{abstract}

Spectral energy distributions (SEDs) from X-ray to far-infrared (FIR) wavelengths are presented for a sample of 1246 X-ray luminous active galactic nuclei (AGN; $L_{0.5-10\ts\rm{keV}}>10^{43}\ts$erg s$^{-1}$), with $z_{\rm{spec}}<1.2$, selected from Stripe 82X, COSMOS, and GOODS-N/S. The rest-frame SEDs show a wide spread ($\sim2.5\ts$dex) in the relative strengths of broad continuum features at X-ray, ultraviolet (UV), mid-infrared (MIR), and FIR wavelengths. A linear correlation (log-log slope of 0.7$\pm0.04$) is found between $L_{\rm{MIR}}$ and $L_{\rm{X}}$. There is significant scatter in the relation between the $L_{\rm{UV}}$ and $L_{\rm{X}}$ due to heavy obscuration, however the most luminous and unobscured AGN show a linear correlation (log-log slope of 0.8$\pm0.06$) in the relation above this scatter. The relation between $L_{\rm{FIR}}$ and $L_{\rm{X}}$ is predominantly flat, but with decreasing dispersion at $L_{\rm{X}}>10^{44}\ts$erg s$^{-1}$. The ratio between the ``galaxy subtracted'' bolometric luminosity and the intrinsic $L_{\rm{X}}$ increases from a factor of $\sim$$10-70$ from log $L_{\rm{bol}}/{\rm(erg\; s}^{-1})=44.5-46.5$. Characteristic SED shapes have been determined by grouping AGN based on relative strengths of the UV and MIR emission. The average $L_{1\mu\rm{m}}$ is constant for the majority of these SED shapes, while AGN with the strongest UV and MIR emission have elevated $L_{1\mu\rm{m}}$, consistent with the AGN emission dominating their SEDs at optical and NIR wavelengths. A strong correlation is found between the SED shape and both the $L_{\rm{X}}$ and $L_{\rm{bol}}$, such that $L_{\rm{bol}}/L_{\rm{X}}=20.4\pm1.8$, independent of the SED shape. This is consistent with an evolutionary scenario of increasing $L_{\rm{bol}}$ with decreasing obscuration as the AGN blows away circumnuclear gas.
\end{abstract}

\keywords{Active galactic nuclei -- AGN host galaxies -- X-ray active galactic nuclei -- Spectral energy distributions -- Surveys}

\section{Introduction} \label{sec:intro}
Nearly every galaxy has a supermassive black hole (SMBH) at its center with a mass ranging from $10^{5}M_\odot$ to $10^{10}M_\odot$. These SMBHs undergo significant periods of growth as active galactic nuclei (AGN), when large amounts of gas and dust accrete onto the SMBH. While many theoretical models require AGN feedback to play a significant role in galaxy evolution in order to match local observations, the exact impact of an AGN on the properties of the host galaxy is still uncertain \citep[see reviews by][]{Alexander2003,Kormendy2013,Heckman2014}. Some studies find suppressed star formation for the most luminous AGN \citep[e.g.,][]{Page2012,Bongiorno2012,Barger2015,Yang2017} and others report that AGN activity does not significantly affect the global star formation or that it may even enhance star formation in some host galaxies \citep[e.g.,][]{Lutz2010,Mullaney2012,Rovilos2012,Santini2012,Suh2019}. To determine the extent of AGN feedback and the effect that this has on the host galaxy, it is vital to disentangle the contributions of the AGN and host galaxy to the total emission. This is challenging as many extragalactic sources contain a mix of star formation and obscured black hole growth. To fully understand the properties of both the AGN and the host galaxy, it is necessary to accurately identify the extent of the emission from the AGN, star formation, and stellar populations in the nearby environments. This can be done through a detailed analysis of their spectral energy distributions (SEDs), with a particular focus on the infrared properties \citep[e.g.,][]{Assef2013,Assef2015,Hickox2018}. If a complete understanding of the intrinsic properties of the AGN and host galaxies is desired, it is also vital in such an analysis, to ensure that each source is properly identified as being dominated by AGN activity, star formation in the host galaxy, or a strong combination of the two. If the AGN emission is not properly accounted for, then derived host galaxy properties, such as the star formation rate (SFR), can be over estimated by up to, if not greater than $\sim$35$\%$ \citep{Kirkpatrick2017,Cooke2020}. 

The accretion of gas and dust onto the SMBH produces a variety of emission properties that are directly tied to the physical characteristics of the AGN, including X-ray emission from the hot corona, ultraviolet (UV)-optical emission directly from the inner accretion disk and mid-infrared (MIR) emission from a warm dusty torus. \citep[for review, see][]{Netzer2015}. Far-infrared (FIR) emission from cold dust may also be directly associated with the AGN or with star formation in the host galaxy \citep[e.g.,][]{Symeonidis2016,McKinney2021}. The obscuration of AGN may be driven by the SMBH accretion properties \citep[][]{Ricci2017b} or by dust located in the torus, polar regions, or at large distances from the galactic nucleus within the host galaxy \citep{Goulding2012,Gilli2022}. Furthermore, there have been studies suggesting that the nuclear dust is not uniformly distributed around the central engine, indicating the complex and clumpy structure of the dusty torus that may include polar dust extending above and below this torus \citep[e.g.,][]{Ramos2009,Ramos2011,Markowitz2014,Ichikawa2015,Asmus2016,Stalevski2017}. \cite{Sanders1988} suggested an evolutionary scenario for AGN in which the obscuration is an evolutionary phase triggered by an accretion event or a merger between two galaxies. Subsequently, the obscured AGN expel most of the obscuring material, evolving into a classic unobscured quasar \citep{DiMatteo2005, Hopkins2006, Ananna2022, Ananna2022b}.

Most known quasars at high redshift were found in optical surveys, primarily in the Sloan Digital Sky Survey \citep[SDSS;][]{Fan2003,Richards2004,Schneider2005}, which was most sensitive to unobscured objects. However, significant periods of growth can also take place in obscured AGN \citep{Hopkins2008} and in fact it is now believed that, the majority of SMBH growth is heavily dust obscured \citep[e.g.,][]{Treister2004,Hickox2007,Treister2009,Ananna2019}. Therefore many AGN undergoing significant periods of growth are missed by classical optical surveys. Selecting AGN by the hard X-ray emission minimizes this bias, as the X-rays that are produced in AGN can pierce through the optically obscuring gas (typically with $N_{\rm H} \sim10^{22}$), so long as it is not Compton-thick, with $N_{\rm H} > 10^{24} \; {\rm cm^{-2}}$ \citep{Bassani1999,Heckman2005}. Powerful X-ray emission seems to be a common feature among nearly all AGN and is produced by the inverse Compton scattering of photons from the accretion disk in the surrounding corona \citep{Nandra1994}. Additionally, purely star-forming galaxies, which produce X-rays through a number of high mass X-ray binaries, young supernova remnants, and hot plasma in star forming regions, rarely reach comparable luminosities to those of powerful AGN in the X-rays, with purely star forming galaxies typically showing $L_{\rm X} < 10^{42}\ts$erg s$^{-1}$ \citep[e.g.,][]{Fabbiano1989,Ranalli2003}. However, X-ray selection may still miss low luminosity or heavily obscured AGN that might instead be detected through their MIR colors \citep{Lacy2004,Lacy2013,Stern2005,Assef2010,Assef2013,Donley2012,Kirkpatrick2017,Hviding2022}. 

To fully understand the energetics and demographics of AGN, it is necessary to characterize their emission across the electromagnetic spectrum, particularly in the MIR and FIR, where much of the energy is emitted but not always observed. To this end, we present an analysis of the SEDs for a sample of $\sim1200$ hard X-ray selected AGN from the Accretion History of AGN (AHA) wedding-cake survey (PI M. Urry). In \S \ref{sec:data} of this paper we describe the multiwavelength data from the four different observation fields used in this analysis. \S \ref{sec: methods} presents the sample selection along with the construction of the X-ray to FIR SEDs. An in-depth analysis of the SED properties with respect to redshift and bolometric luminosity is presented in \S \ref{sec:analysis}. \S \ref{sec:discussion} presents a breakdown of the characteristic SED profiles and presents the emission properties in the context of the AGN life-cycle. Finally, our results are summarized in \S \ref{sec:summary}. Throughout this analysis we assume a standard cosmology with $H_{0} = 70$\,km s$^{-1}$ Mpc$^{-1}$, $\Omega_{m} = 0.3$, and $\Omega_{\Lambda} = 0.7$.

\section{Data} \label{sec:data}
This work utilizes detailed multiwavelength data from the AHA survey, which is comprised of three survey fields of varying depths and sky coverage. Together, these fields create a ``wedding-cake survey,'' where each field probes a different area and flux limit, and therefore detects AGN in a different luminosity and redshift range. The widest field is Stripe~82X \citep{LaMassa2013_Chandra,LaMassa2013_XmmNewton_Chandra,LaMassa2016,Ananna2017}, a wide-area X-ray survey covering $\sim31 \; \rm{deg}^{2}$ of the legacy SDSS Stripe 82 field \citep{Jiang2014}. The middle layer of the ``wedding-cake'' is the COSMOS field \citep{Scoville2007, Elvis2009, Civano2016}, a deep, wide field, multi-wavelength survey covering a 2 $\rm{deg}^{2}$ field centered on the J2000 coordinates, $RA = +150.119$ and $Dec = +2.205$. The final layer is from the Great Observatories Origins Survey \citep[GOODS;][]{Giavalisco2004}. The two GOODS fields, GOODS North and South (GOODS-N/S), utilize extremely deep observations from NASA's Great Observatories; together these two fields cover a total of 320 square arcminutes, and contain the deepest flux limits in the AHA survey.

\subsection{Stripe~82X} \label{sec:s82x}
Stripe~82X contains three contiguous regions of XMM-{\it Newton} coverage, observed in XMM-{\it Newton} cycles 10 and 13 (AO10 and AO13) along with archival XMM-{\it Newton} and {\it Chandra} data. The details of the X-ray observations are described in \cite{LaMassa2013_Chandra, LaMassa2013_XmmNewton_Chandra,LaMassa2016}, with a final multiwavelength catalog and photometric redshifts presented by \cite{Ananna2017}. Stripe~82X contains 6181 X-ray sources. Of these, \cite{Peca2022} identified 2937 AGN with reliable redshifts and sufficient counts to perform X-ray spectral modeling to determine the intrinsic X-ray luminosity that is corrected for galactic, host galaxy and circumnuclear extinction along with estimates of the neutral hydrogen column density ($N_{\mathrm{H}}$) for these sources. 

In addition to the X-ray coverage, Stripe~82X also contains rich multiwavelength data with ultraviolet (UV) data from GALEX \citep{Morrissey2007}, optical data from deep co-added SDSS Stripe 82 catalogs \citep{Jiang2014,Fliri_Trujillo2016}, near-infrared (NIR) data from UKIDS \citep{Hewett2006,Casali2007,Lawrence2007} and the Vista Hemisphere Survey \citep[VHS;][]{McMahon2013}, mid-infrared (MIR) from {\it Spitzer} IRAC \citep{Timlin2016,Papovich2016} and the {\it Wide-field Infrared Survey Explorer} \citep[WISE;][]{Wright2010}, far-infrared (FIR) coverage from {\it Herschel} Spire \citep{Viero2014}, and radio coverage at 1.4 GHz from FIRST \citep{Becker1995,Helfand2015}. 

\cite{Ananna2017} utilized these data to construct the latest version of the Stripe~82X multiwavelength catalog. The various multiwavelength associations were cross-matched using a statistical maximum likelihood estimator algorithm to report the most likely counterpart to each X-ray source \citep{Sutherland1992}, allowing for much more accurate cross-matches than a simple positional matching technique. The final catalog consists of far- and near-ultraviolet (FUV and NUV) from GALEX (0.15 and 0.23$\,\mu$m respectively); u, g, r, i, and z, from SDSS (0.34, 0.48, 0.62, 0.77, 1.1$\,\mu$m respectively); J, H, and K from VHS/UKIDSS (1.25, 1.63, 2.20$\,\mu$m respectively); CH1 and CH2 from IRAC (3.54 and 4.48$\,\mu$m respectively); and W1, W2, W3, and W4 from ALLWISE (3.35, 4.6, 11.6, and 22.1$\,\mu$m). Finally, the matches to the FIR {\it Herschel} SPIRE data (250, 350, and 500$\,\mu$m) which were not included in \cite{Ananna2017}, are taken from \cite{LaMassa2016} based on the matching ID of each source. In this work we substitute the ALLWISE W3 and W4 data reported in \cite{Ananna2017} with that from \cite{Lang2016}, who utilized a forced photometry technique, using measured SDSS source positions, star-galaxy classifications, and galaxy profiles to define the sources whose fluxes are to be measured in the WISE images. This results in a greater number of detections in the W3 and W4 bands and more sensitive flux limits.

\subsection{COSMOS} \label{sec:cosmos} 
The COSMOS-Legacy survey is a 4.6 Ms {\it Chandra} program that imaged the COSMOS field. Details of this survey, including X-ray and optical/infrared photometric and spectroscopic properties, are described in \cite{Civano2016} and \cite{Marchesi2016}. \cite{Marchesi2016b} and \cite{Lanzuisi2018} conducted an extended analysis of {\it Chandra} COSMOS Legacy sources with more than 30 net counts, utilizing improved background modeling and modeling techniques optimized for heavily obscured sources. These catalogs were built off the initial work done in the XMM-COSMOS survey \citep{Cappelluti2007,Cappelluti2009}.

\cite{Weaver2021} presented the updated catalog of precise photometric redshifts and 30-band photometry for more than half a million secure objects within the 2 deg$^{2}$ COSMOS field. This updated catalog presents significantly deeper optical and NIR images along with a reprocessing of all {\it Spitzer} data. The aperture extraction techniques described in \cite{Weaver2021} ensure that accurate data are presented for each photometric filter, with a low probability of contamination from differing sources, particularly in the UV to MIR wavelengths. \cite{Marchesi2016} provides the cross-matches between the COSMOS-Legacy survey and the original COSMOS mulitwavelength catalog from \cite{Laigle2016}. The IDs from this cross match were used to identify the mulitwavelength counterparts in the \cite{Weaver2021} catalog, to the X-ray sources from \cite{Marchesi2016}, \cite{Marchesi2016b}, and \cite{Lanzuisi2018} as well as the FIR data that was presented in \cite{Laigle2016}, but absent from the updated catalog. The final catalog includes the GALEX FUV and NUV band; CFHT U band; five Subaru Hyper SuprimeCam (HSC) bands (g, r, i, z, y); four UltraVista bands (Y, H, J, Ks); four {\it Spitzer} IRAC bands (3.6, 4.5, 5.8, and 8.0$\,\mu$m); the 24$\,\mu$m {\it Spitzer} MIPS observations; and the {\it Herschel} PACS (100$\,\mu$m and 160$\,\mu$m) and SPIRE (250$\,\mu$m, 350$\,\mu$m, and 500$\,\mu$m) observations. The details of these mulitwavelength observations can be found in \cite{Weaver2021} and references within. Additional spectroscopic redshifts were also gathered from \cite{Hasinger2018} and matched to IDs from the COSMOS catalogs. 

Significant efforts have previously been made to analyze the SEDs of X-ray luminous AGN in the COSMOS field, with a particular focus on unobscured AGN and determining their characteristic multiwavelength features \citep[e.g.][]{Elvis2012,Hao2013,Hao2014} or comparing the properties and populations statistics of obscured and unobscured AGN \citep[e.g.,][]{Lusso2013,Suh2019}. This analysis builds off of these previous works by placing the COSMOS field in the greater context of the entire AHA wedding cake survey.

\subsection{GOODS} \label{sec:GOODS}
The GOODS-N and GOODS-S fields are centered around the {\it Chandra} Deep Field (CDF) surveys \citep[see, e.g.,][ for reviews]{Brandt2015,Xue2017}. These two fields consist of the 7 Ms {\it Chandra} Deep Field South \citep[CDF-S;][]{Luo2017}, which is the deepest X-ray survey to date, and the 2 Ms {\it Chandra} Deep Field North \citep[CDF-N;][]{Xue2016}. These deep X-ray surveys allow for a detailed analysis of the X-ray spectra. With these observations, accurate estimates of $N_{\mathrm{H}}$ and intrinsic X-ray luminosity can be made. \cite{Li2020} improved this analysis for the most heavily obscured AGN within CDF-N/S. Estimates of the intrinsic X-ray luminosity and column density are taken from \cite{Li2020} when available.

GOODS-N/S comprise two of the five Cosmic Assembly Near-infrared Deep Extragalactic Legacy Survey \citep[CANDELS;][]{Grogin2011,Koekemoer2011} fields. CANDELS is a {\it Hubble Space Telescope} (HST) 902-orbit legacy program designed to study galaxy formation and evolution over a wide range of redshifts using the NIR HST/WFC3 camera to obtain deep imaging of faint and distant objects. \cite{Guo2013} presented a catalog of the UV to MIR data for the CANDELS/GOODS-S filed by combining the HST data (ACS: F435W, F606W, F775W, F814W, F850LP, and WFC3 F098M) with archival data from CTIO/MOSAIC, VLT/VIMOS, VLT/ISAAC $K_{s}$, VLT/HAWK-I $K_{s}$, and {\it Spitzer}/IRAC 3.6, 4.5, 5.8, 8.0$\,\mu$m. The {\it Spitzer}/MIPS 24 and 70$\,\mu$m and {\it Herschel} 100, 160, 250, 350, and 500$\,\mu$m data come from matches to \cite{Elbaz2011}.

\cite{Barro2019} presented a catalog of the CANDELS/GOODS-N field containing the photometry from the UV to FIR. This includes data from the U band from KPNO and LBC, HST/ACS F435W, F606W, F775W, F814W and F850LP, HST/WFC3 F105W, F125W, F140W, and F160W, Subaru/MOIRCS Ks; CFHT/MegaCam K; and {\it Spitzer}/IRAC 3.6, 4.5, 5.8, and 8.0 $\,\mu$m and FIR {\it Spitzer}/MIPS 24$\,\mu$m {\it Herschel}/PACS 100 and 160$\,\mu$m, SPIRE 250, 350, 500 $\mu$m. 

These two catalogs were combined with additional data in the optical to MIR gathered from \cite{Hsu2014}, \cite{Yang2014}, \cite{Skelton2014}, \cite{Damen2011}, \cite{Straatman2016}  along with data in the FIR from \cite{Oliver2012} and the ``Super-deblended'' FIR data presented in \cite{Liu2018}. This extensive list of multiwavelength sources for the GOODS fields was matched using a nearest neighbor approach, utilizing a matching radius of 2\arcsec\ for the UV-NIR data and a matching radius of 5\arcsec\ for the MIR-FIR data. These matches were then confirmed by comparing the redshifts reported in each respective catalog when available.

For all three surveys, significant work has been previously done to ensure a low probability of sources being incorrectly matched across different photometric filters, either through the use of maximum likelihood matching techniques \citep{Ananna2017}, carefully defined aperture extraction \citep{Weaver2021}, or positional matching with narrow matching radii plus comparison of spectroscopic information, such as redshift. The {\it Herschel} SPIRE FIR images are the most likely source of mismatches, due to the lower spatial resolution of the {\it Herschel} SPIRE instrument relative to other wavelengths. To check for contamination in SPIRE-detected sources, we have compared the source density of the higher resolution MIR detections from the {\it Spitzer} MIPS $24\ts\mu$m data to the beam size of the {\it Herschel} instrument. We find that, on average, only one source with a large enough flux to strongly contribute to the reported FIR flux of the sources in the sample is found within the typical beam size of the {\it Herschel} SPIRE data. This is consistent with deblending work in the COSMOS field reported by \cite{Jin2018}. Therefore, while contamination may play a role in some sources, it does not affect the majority of the sample ($<10\%$).

\section{Methods} \label{sec: methods}
\subsection{Sample Selection} \label{sec: target selection}

As the goal of this work is to comprehensively analyze the SEDs of luminous AGN from the AHA fields, we limit our sample to those sources with photometric coverage in the regions of the electromagnetic spectrum necessary for such an analysis, namely the UV and MIR region where the emission from the accretion disk and dusty torus respectively is most prominent. Each of the survey fields has extensive coverage in the UV and optical, therefore having adequate data to analyze the accretion disk emission will not be an issue for a majority of sources. However, the MIR coverage for these sources is much more sparse, with large gaps between the observed filters. As discussed in \S \ref{sec:data}, COSMOS and GOODS-N/S have MIR data in the {\it Spitzer} IRAC channels going out to $8\,\mu$m, the {\it Spitzer} MIPS data at $24\,\mu$m, before a large gap going to the FIR data. Stripe~82X has similar coverage with the WISE W3 and W4 bands at $11\,\mu$m and $22\,\mu$m. This poses a larger challenge for analyzing the MIR data of AGN at high redshift. At $z > 1.2$, the {\it Spitzer} MIPS $24\,\mu$m data and the WISE W4 data are shifted to rest-frame wavelengths blue-ward of $10.9\,\mu$m, and $10\,\mu$m, respectively. For this reason, we limit our sample to sources in the AHA fields with a secure spectroscopic redshift measurement less than 1.2, ensuring that we have a photometric observation in the rest-frame MIR near 10$\,\mu$m for all sources in our sample. This will allow us to better characterize the MIR SED profiles and investigate how the 10$\,\mu$m traces AGN properties. 

\begin{figure}[t!]
    \centering
    \includegraphics[width=\linewidth]{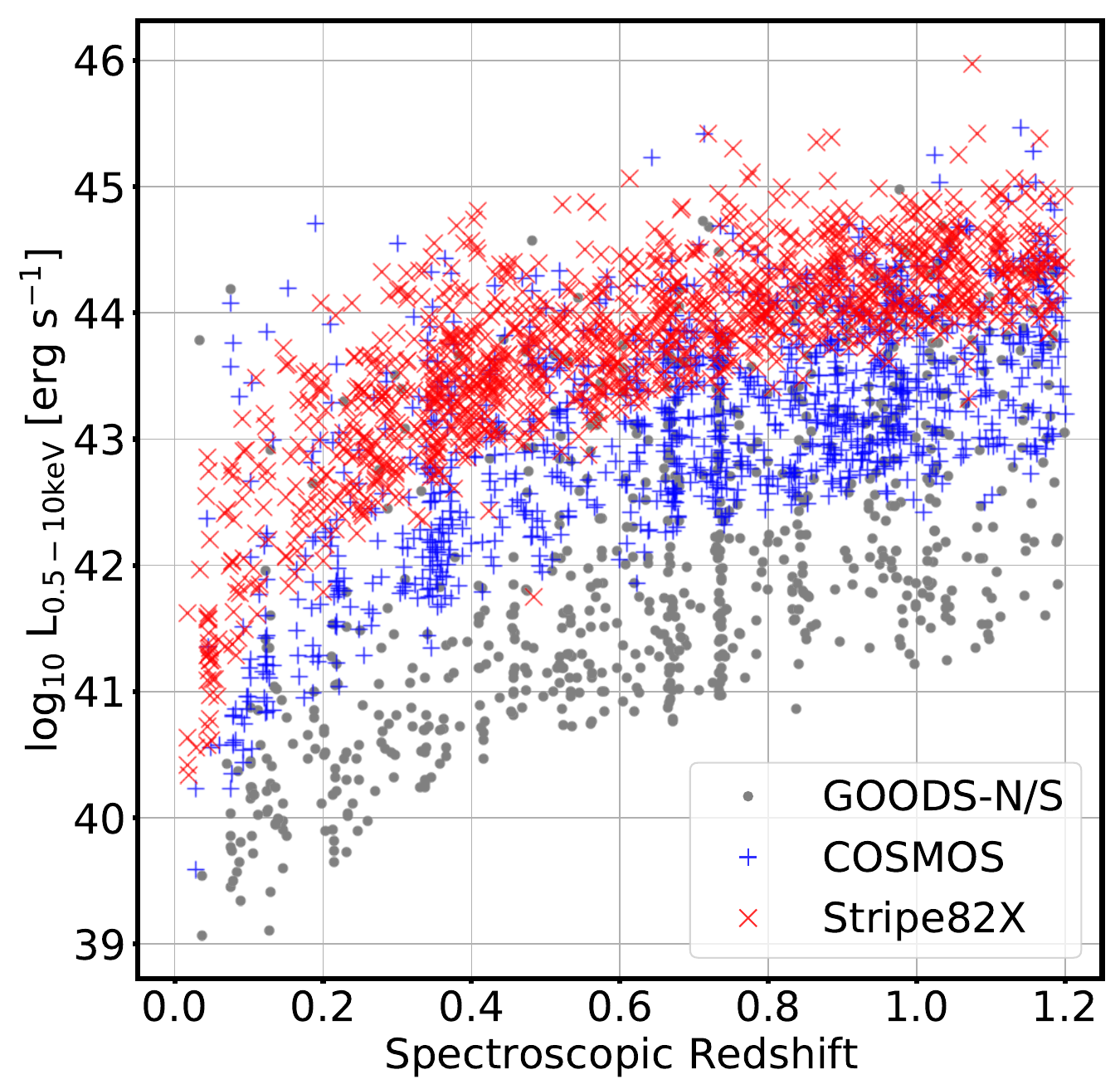}
    \caption{The intrinsic 0.5--10$\ts$keV X-ray luminosity as a function of spectroscopic redshift for the sources in the AHA survey. The red x symbols are the sources from largest and shallowest layer of the wedding cake, the Stripe~82X survey, the blue plus symbols are from the middle layer, COSMOS and the gray points are from the GOODS-N/S fields. Only sources with confirmed multiwavelength counterparts that have been observed with {\it Herschel} are shown.}
    \label{fig:Lx_z_scatter}
\end{figure}

In order to analyze the cold dust emission, we further limit the sample by removing sources that were not within the fields covered in the FIR with the {\it Herschel} SPIRE instrument (250, 350, and 500$\,\mu$m). The entire COSMOS field and both GOODS fields were covered with SPIRE, however only a portion of Stripe~82X was observed in these bands. The X-ray fields within Stripe~82 that fall in an R.A. range of 13$^{\circ}$ to 37$^{\circ}$ and a declination of $-2^{\circ}$ to 2$^{\circ}$ are included in this analysis. This is the full XMM-AO13 area, as the fields for these observation were specifically chosen to overlap with the {\it Herschel} and {\it Spitzer} coverage within Stripe 82. Sources that do not fall in these regions are removed from the sample so upper limits can be used to accurately constrain the FIR emission.

\begin{deluxetable}{cccc}
\tablecaption{The number of sources from each AHA field included in the analysis.}
\tablehead{
\colhead{Field} & \colhead{$z < 1.2$} & \colhead{$L_{\mathrm{X}} >$ 10$^{43}$} & \colhead{Rest-Frame 6$\,\mu$m} \\  & & \colhead{[erg s$^{-1}$]} & 
}
\colnumbers
\startdata
         Stripe~82X & 1169 & 998 & 529 \\
         COSMOS & 1128 & 664 & 624 \\
         GOODS$-$N/S & 704 & 152 & 93 \\
         \hline
         Total & 3001 & 1814 & 1246 \\
\enddata
\tablecomments{(1) The AHA field. (2) Number of sources from each field with a spectroscopic redshift of $z \leq 1.2$ and are located in a field observed by {\it Herschel} SPIRE. (3) The number of sources from column 2 that satisfy the minimum intrinsic 0.5--10\,kev X-ray luminosity condition of $L_{\mathrm{X}} >$ 10$^{43}$ erg s$^{-1}$ for the sample selection. (4) The number of these sources that have secure mulitwavelength counterparts and detections near rest-frame $\sim$6$\,\mu$m.}
\label{tab:SourceCount}
\end{deluxetable}

Figure \ref{fig:Lx_z_scatter} shows the intrinsic X-ray luminosity as a function of spectroscopic redshift for all sources with $z < 1.2$ and observations (though not necessarily detections) with {\it Herschel} SPIRE. This figure also illustrates the benefits of a wedding cake style X-ray survey by clearly showing the effects of the different flux limits and area of the different fields, with Stripe~82X capturing the most luminous sources, thanks to the increased volume accessed by the wide survey field. GOODS-N/S captures the least luminous sources, thanks to the increased depth. The COSMOS field then bridges the gap between the two. It can be seen that the combined surveys are substantially complete at X-ray luminosites greater than 10$^{42}$ erg s$^{-1}$ up to $z\sim$1.2.

AGN are identified from this $z < 1.2$ sub-sample based on their intrinsic 0.5--10 keV X-ray luminosity taken from \cite{Peca2022}, \cite{Ananna2017}, \cite{Marchesi2016}, \cite{Marchesi2016b}, \cite{Lanzuisi2018}, \cite{Xue2017}, \cite{Luo2017}, and \cite{Li2020}. AGN can generate X-ray luminosities up to $10^{46}$ erg s$^{-1}$ while star forming galaxies rarely produce X-ray luminosities greater than $10^{42}$ erg s$^{-1}$ \citep{Ranalli2003,Persic2004}. For this sample, all sources with intrinsic X-ray luminosities in the 0.5--10 keV band with $L_{\rm{X}} < 10^{43}$ erg s$^{-1}$ are removed from the sample to prevent contamination by even the most luminous purely star forming galaxies. While this limit likely removes some low luminosity AGN, it will ensure a clean sample of purely luminous AGN and little to no contamination from non-AGN sources. This is also where the sample is most complete, allowing us to include AGN that are luminous in the X-rays, even though the non-AGN emission may dominate at other wavelengths \citep{Brandt2015}.

\begin{figure}
    \centering
    \includegraphics[width=\linewidth]{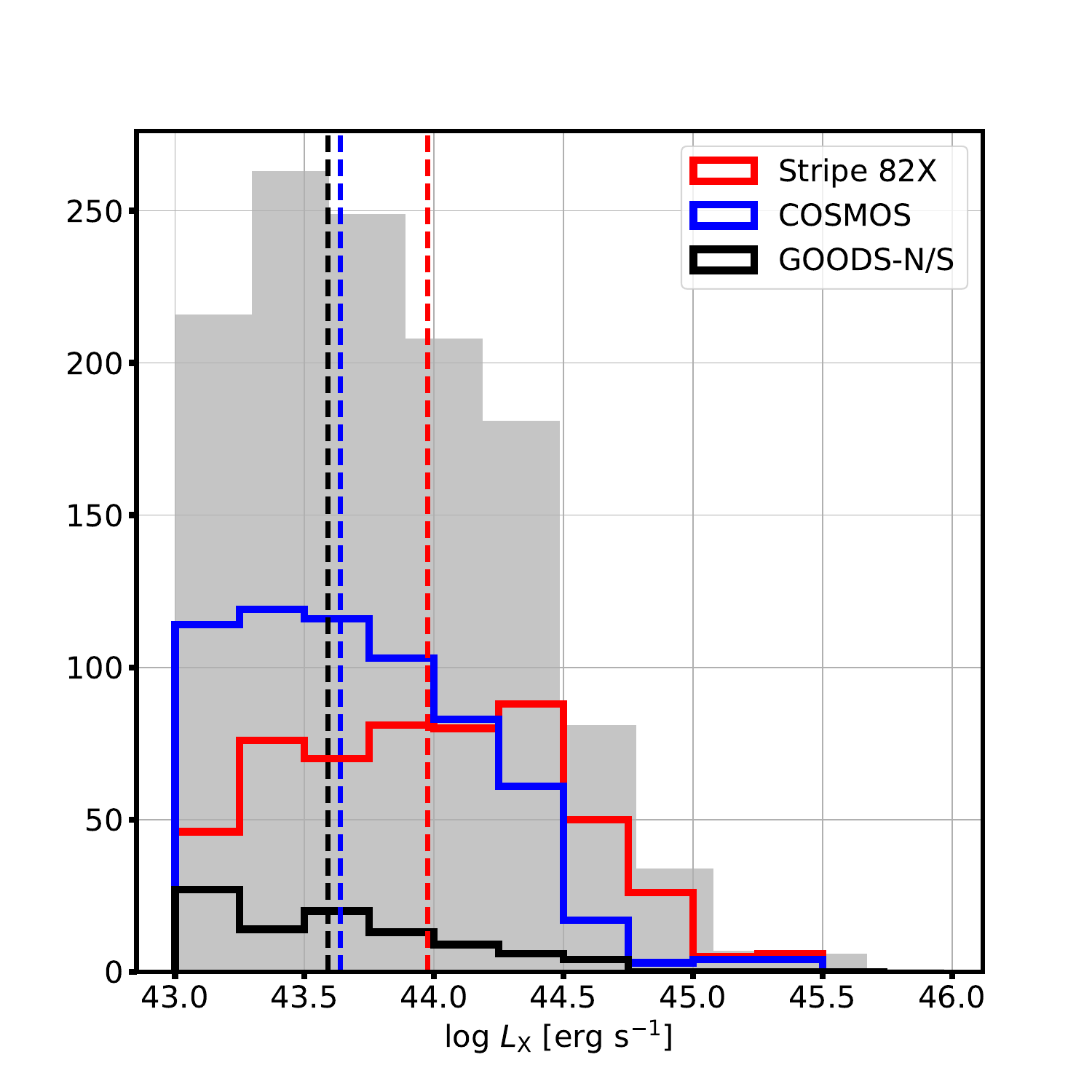}
    \caption{Histogram of the intrinsic 0.5--10$\ts$keV X-ray luminosity for the 1246 sources in our final sample. The gray histogram is the total sample, the histogram of the sources from each AHA field (Stripe~82X - red, COSMOS - blue, GOODS-N/s - black) are also shown. The median of each distribution from the respective AHA fields is shown as a vertical dashed line in the same corresponding color.}
    \label{fig:Lx_hist_sample}
\end{figure}

 \begin{figure*}
    \centering
    \includegraphics[width=\textwidth]{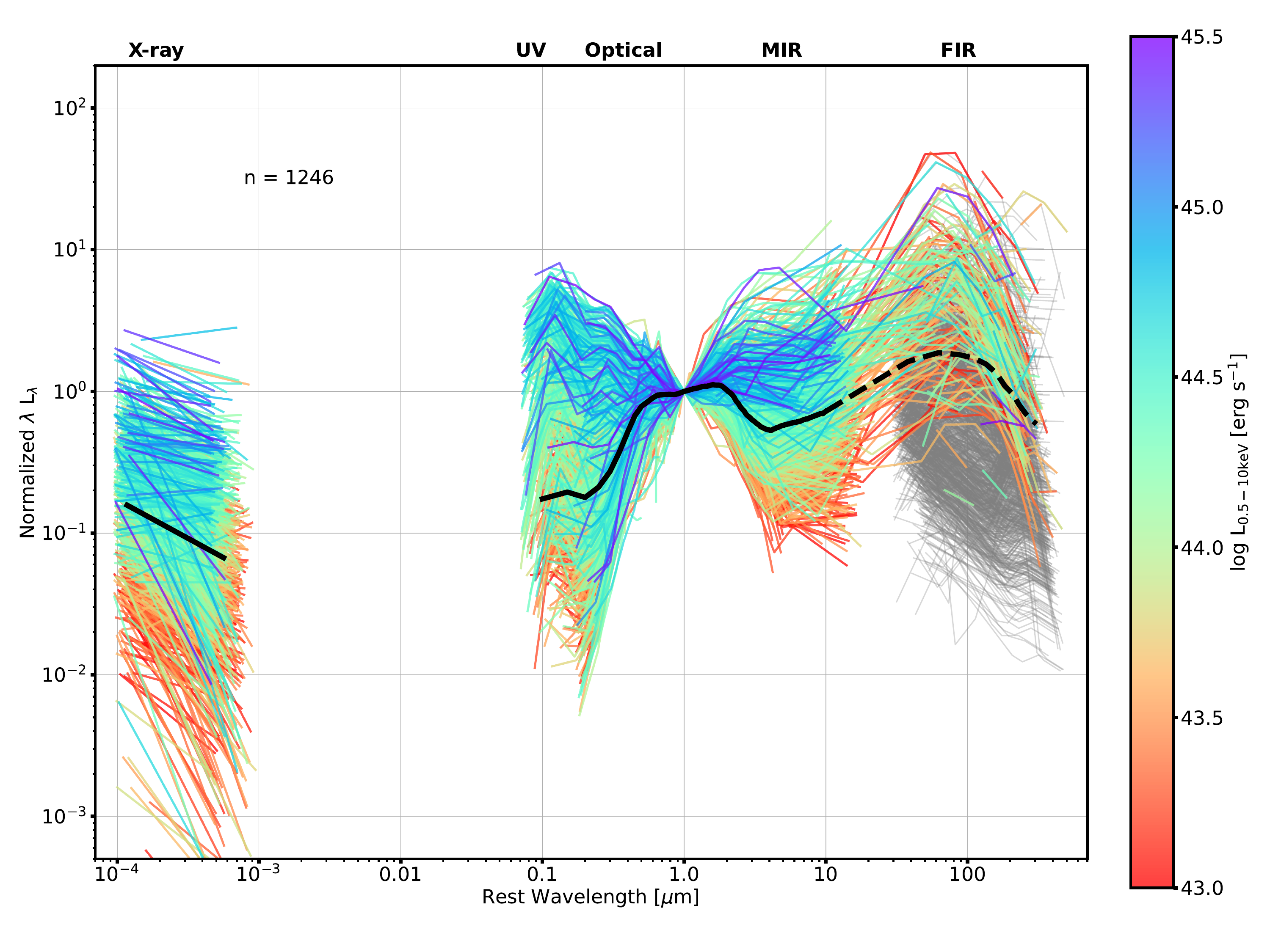}
    \caption{The SEDs normalized at 1$\,\mu$m for the sources within our sample that match the criteria outlined in \S \ref{sec: target selection} and Table \ref{tab:SourceCount}. The color of each SED is the intrinsic 0.5--10 keV X-ray luminosity. The solid black line is interpolation of the median SED in the X-ray and in the UV-MIR region. The dashed black line is the interpolation of the median SED of the detections and the 1$\sigma$ upper limits in the FIR normalized at 1$\,\mu$m. The normalized FIR upper limits for individual sources are shown in gray. A wide range of emission properties in the X-ray, UV, MIR, and FIR can be seen around the total median SED.}
    \label{fig:tot_SEDs}
\end{figure*}

\begin{figure}
    \centering
    \includegraphics[width=\linewidth]{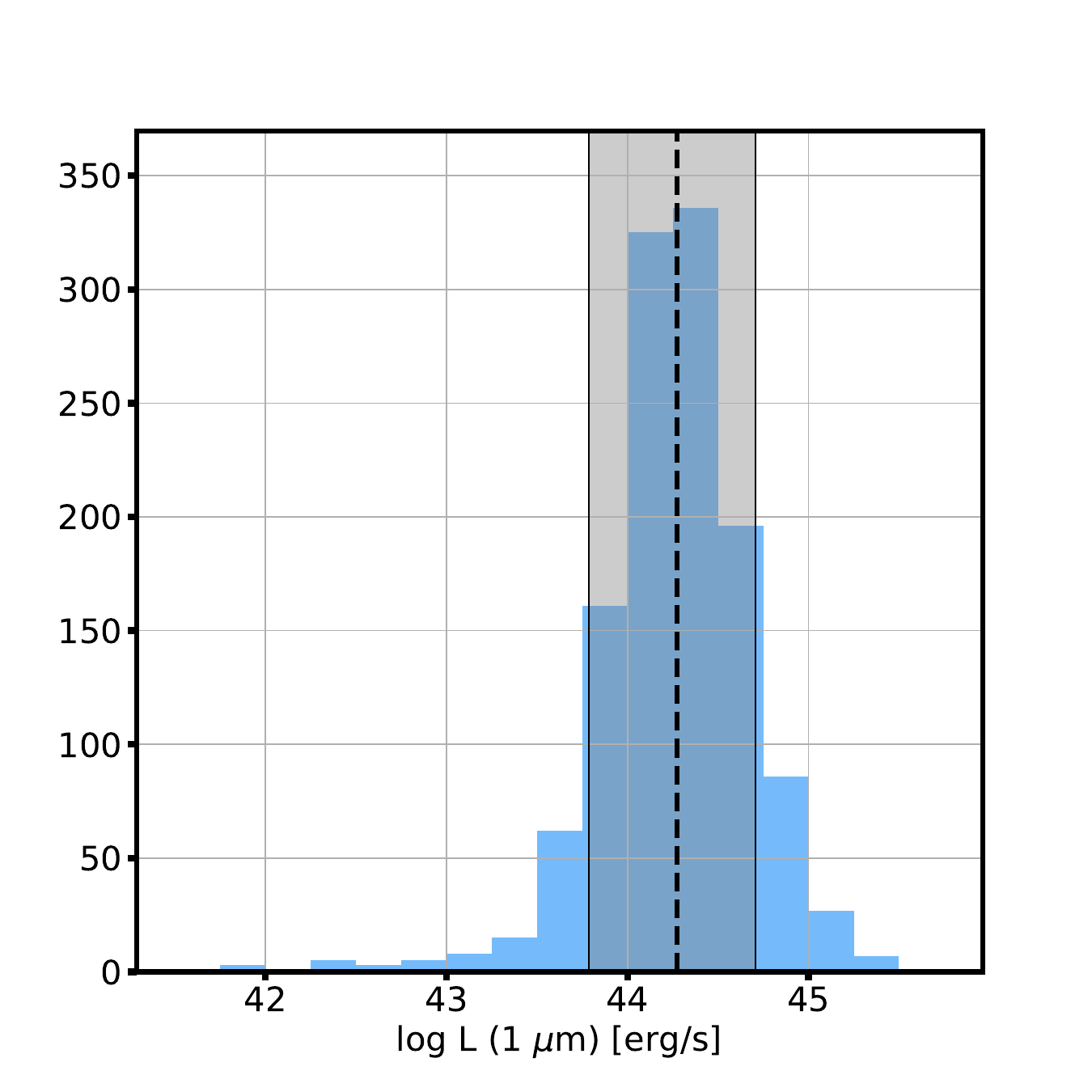}
    \caption{A distribution of the 1$\,\mu$m luminosities for the 1246 AGN in our sample. The mean of the distribution is shown as a vertical dashed line and the shaded region shows the 1$\sigma$ spread about the average. The 1$\sigma$ distribution spans $\pm$0.46 dex around the mean of log $L (1\mu \rm m)/{\rm (erg~s}^{-1})$ = 44.2.}
    \label{fig:Lone_hist}
\end{figure}

\begin{figure*}
    \centering
    \includegraphics[width=\textwidth]{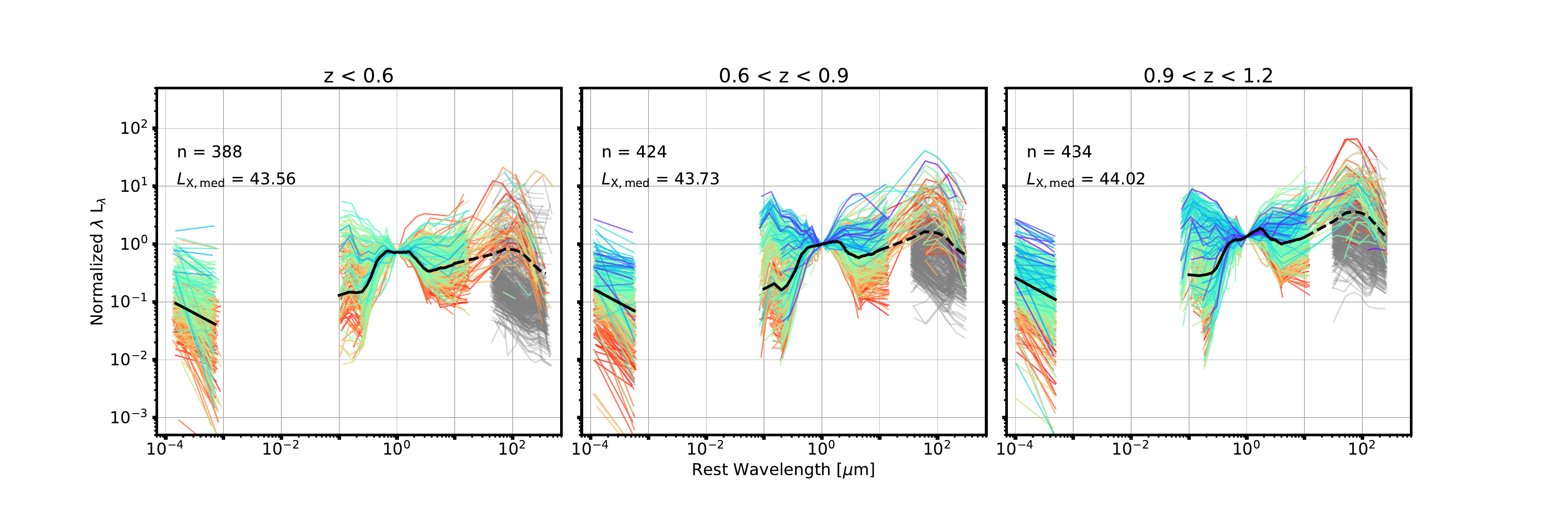}
    \caption{SEDs separated into three redshift bins. Each SED is normalized at 1$\,\mu$m with the normalization of each bin scaled to the normalization of the central bin. The color of each SED is the intrinsic 0.5--10 keV X-ray luminosity with the same colorbar found in Figure \ref{fig:tot_SEDs}. The solid black line is the median SED for the sources in each bin. The dashed black line is the median of the detections, stacked fluxes, and the 1$\sigma$ upper limits in the FIR. The FIR upper limits for individual sources are shown in gray. No major change in the median SED can be seen between the three redshift bins.} 
    \label{fig:SEDs_zbins}
\end{figure*}

\begin{figure}
    \centering
    \includegraphics[width=\linewidth]{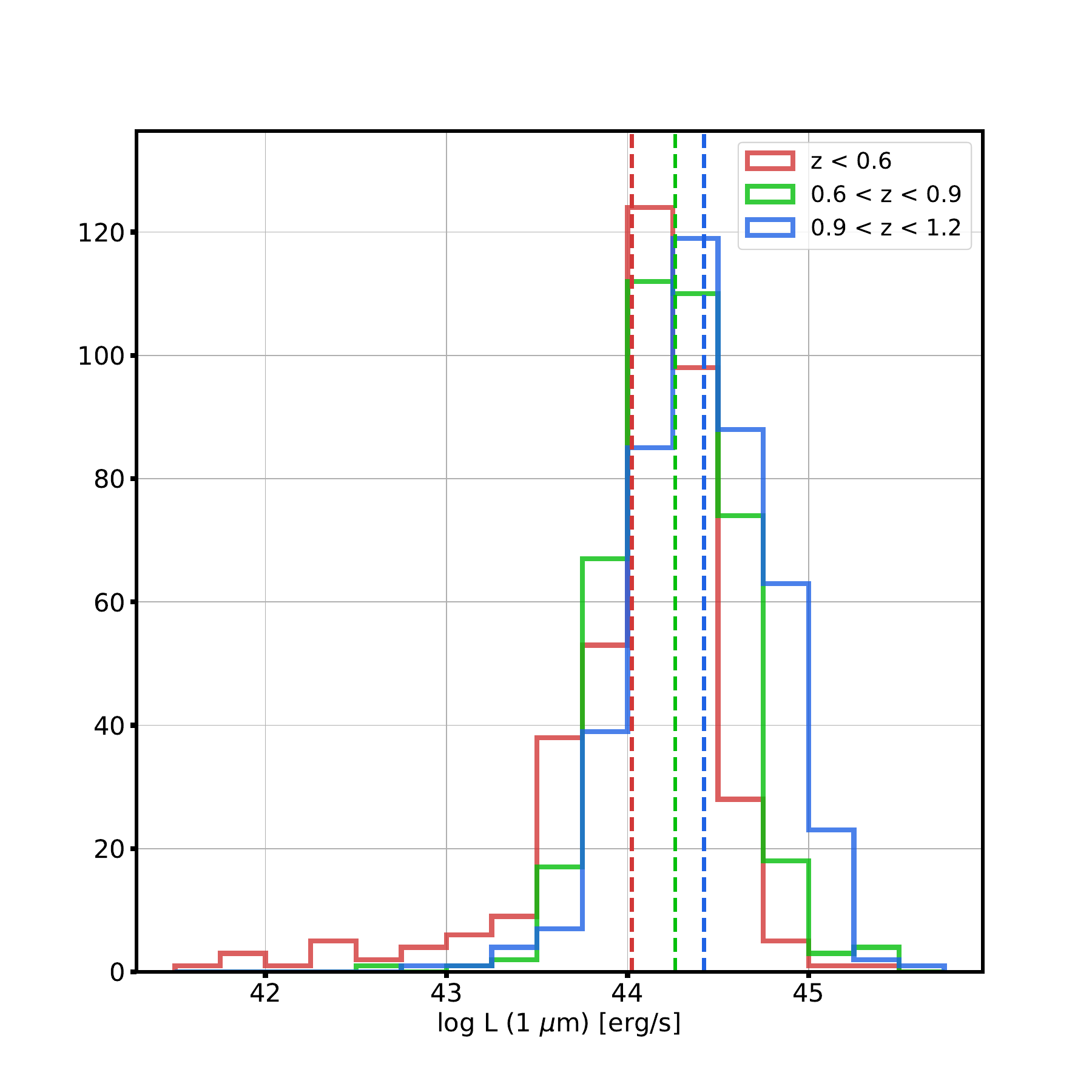}
    \caption{Distribution of the 1$\,\mu$m luminosities for the 1246 AGN in our sample separated into three redshift bins. The mean of each distribution is shown as a vertical dashed line with the same corresponding color as the histogram (log $L_{1\,\mu \rm m}$/(erg$\ts$s$^{-1}$) = 44.02, 44.27, 44.42 for each redshift bin in increasing order).}
    \label{fig:Lone_hist_zbins}
\end{figure}

\begin{figure*}
    \centering
    \includegraphics[width=\textwidth]{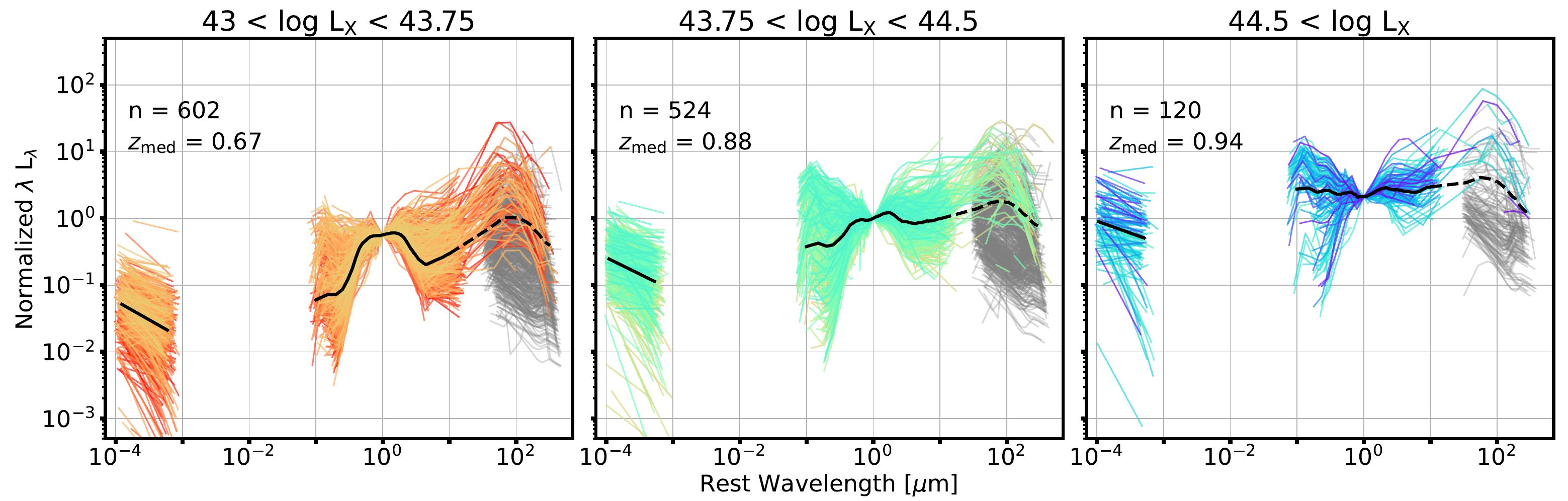}
    \caption{SEDs separated into three bins of increasing intrinsic 0.5--10 keV X-ray luminosity. Each SED is normalized at 1$\,\mu$m with the normalization of other bins scaled to the normalization of the central bin. The color of each SED is the intrinsic 0.5--10 keV X-ray luminosity with the same colorbar found in Figure \ref{fig:tot_SEDs}. The solid black line is the median SED for the sources in each bin. The dashed black line is the median of the detections and the 1$\sigma$ upper limits. The FIR upper limits for individual sources are shown in gray. An evolution in the median SED can be seen with $L_{\rm X}$, however a large spread in the range of each emission property is still visible in each bin.} 
    \label{fig:SEDs_Lxbins}
\end{figure*}

Finally, sources are removed from the sample if they do not have a photometric detection within $\pm2\ts\mu$m of 6$\ts\mu$m in the rest-frame. 6$\ts\mu$m is a key AGN diagnostic wavelength that has been shown to correlate strongly with X-ray properties. In order to have conduct a thorough analysis of AGN properties in the MIR, while being able to compare to previous work, we include this additional requirement for the sample. 

Table \ref{tab:SourceCount} shows the contribution to the total sample of sources from each of the AHA fields. About half of the total sample comes from the COSMOS field, which has the most complete photometric and spectroscopic data set at the X-ray luminosities we are analyzing. Many GOODS sources fall below the X-ray luminosity limit, as can be seen in Figure \ref{fig:Lx_z_scatter}. Additionally, many of the sources from Stripe~82X were excluded due to a lack of detections with low error in the MIR WISE data near a rest frame of $6\ts\mu$m. This requirement of MIR data may preferentially remove most dust poor systems from the sample, however, sources with similar emission properties as those removed from Stripe~82X are found in the COSMOS and GOODS-N/S fields due to the greater sensitivity of {\it Spitzer} in these fields. We additionally find similar distributions in the intrinsic X-ray luminosity for the sources within the final sample as those that are removed, therefore they are likely sampling the same population of AGN. 

Figure \ref{fig:Lx_hist_sample} shows the distribution of intrinsic X-ray luminosity for the final sample of 1246 AGN. Similar to the distribution of the total AHA sample seen in Figure \ref{fig:Lx_z_scatter}, the AGN from Stripe~82X have a larger median X-ray luminosity, with the AGN from GOODS-N/S showing the lowest median X-ray luminosity. 

\subsection{Constructing the SEDs} \label{sec: constructing seds}
The SEDs are constructed by combining the X-ray to FIR data for all sources with each photometric data point corrected to the rest-frame using the known spectroscopic redshift. Each flux measurement is also converted to a luminosity in erg s$^{-1}$ using the same spectroscopic redshift and then normalized by the 1$\,\mu$m luminosity. As the 1$\,\mu$m emission is likely dominated by stellar light from the host galaxy, with little contamination from all but the most luminous AGN \citep{Neugebauer1989}, normalizing the SED at this wavelength allows for the expected features of the AGN emission in the UV and MIR to be easily identified (i.e., the emission from the accretion disk and dusty torus). Contributions of AGN emission at 1$\,\mu$m and how this affects the normalization of the sample are discussed in \S\ref{sec:5 panel seds}. 

In order to simplify the process of normalizing all sources over a broad redshift range at the same rest-frame luminosity, the SED is generated for every source through a simple logarithmic interpolation between each secure photometric data point available for that source. The 1$\,\mu$m luminosity is then taken from this interpolated SED. This same process is followed for the luminosities at other specified wavelengths that are analyzed throughout this work. While individual sources may not have a detection in a filter that falls at the specified wavelength when brought to the rest-frame, each source has secure photometric detections (or upper limits) at observational filters that fall short-ward and long-ward of each specified wavelength, allowing for an estimation of the luminosity at the specified wavelength to be made. If there are no data, or large gaps in the data, short-ward or long-ward of a specified luminosity, then a linear extrapolation is used. 

All 1246 AGN are shown in Figure \ref{fig:tot_SEDs} colored by the intrinsic 0.5--10$\ts$keV X-ray luminosity, with the median SED shown in black. Figure \ref{fig:Lone_hist} shows the distribution of the 1$\,\mu$m luminosities used for the normalization. The standard deviation of the distribution is under an order of magnitude. As stellar emission peaks near 1$\ts\mu$m, this luminosity can be used as a proxy for the stellar mass of these sources, assuming that no AGN emission is strongly contaminating the emission at this wavelength. While this may be a safe assumption for lower luminosity AGN, \cite{Sanders1989} shows that the outer regions of the accretion disk (0.1 -- 1$\ts$pc) can emit strongly between $0.5 - 5 \ts \mu$m through thermal emission. This potential contribution from the AGN near $1\ts\mu$m is examined more closely in \S \ref{sec:5 panel seds}.

While there are adequate detections from X-ray to MIR to accurately interpolate the SED for each source, $\sim70\%$ of the sample is not detected in any of the three {\it Herschel} SPIRE bands. However, by design, each source in the sample was observed in the FIR with {\it Herschel}. This allows us to utilize image stacking to place better constraints on the flux from the FIR. We bin sources by redshift and MIR luminosity and stack the 250$\ts\mu$m images for each bin at the locations of the optical coordinates for each source. This stacking procedure is described in more detail in Appendix \ref{appendix: stacking}. While the image stacking was utilized for the Stripe~82X sources to place improved constraints on FIR emission, this process was unable to be performed on the sources from COSMOS or GOODS-N/S, as no statistically significant signal above the background noise could be found even in the stacked images. Therefore we utilize the 1$\sigma$ upper limits in the {\it Herschel} SPIRE bands for the sources from COSMOS and GOODS-N/S and the flux measurements from the stacked sources in Stripe~82X for the sources undetected in the FIR throughout the rest of the analysis. The 1$\sigma$ upper limits were chosen as a conservative limit so as to not over estimate the FIR luminosity for these sources, whose emission could not be detected, even when stacked. These upper limits were also compared to the nearby galaxy NGC 518. We found the ratio of the rest-frame 100$\ts\mu$m emission to the 1$\ts\mu$m emission comparable to that found using the upper limits in our sample, showing that the FIR luminosity is not drastically underestimated when utilizing 1$\sigma$ upper limits. The dashed black line in Figure \ref{fig:tot_SEDs} is the median of the detections and the 1$\sigma$ upper limits for each source renormalized by the 1$\ts\mu$m luminosity.

\section{Analysis} \label{sec:analysis}

\subsection{SED Evolution with Redshift} \label{sec: Redshift}

In order to account for any potential change in the SED shape due to evolution with redshift, the sample of 1246 AGN is separated into three redshift bins: $0.0 < z < 0.6$ (388 sources), $0.6 < z < 0.9$ (424 sources), and $0.9 < z < 1.2$ (434 sources). These three bins were defined so each contains a roughly equal number of sources and is large enough for a statistical analysis of the multiwavelength properties while still allowing for insight into the evolution of these sources across cosmic time. The normalized SEDs for these three redshift bins can be seen in Figure \ref{fig:SEDs_zbins}. The median SED is largely the same for all three redshift bins, though a slight increase in the relative amount of NIR emission at $\sim$2$\,\mu$m of roughly $\sim$0.1 dex can be seen from the lowest to highest redshift bin. Additionally, the average FIR luminosity increases with redshift by $\sim$0.35 dex. This is to be expected as the dust content of star-forming galaxies increases with redshift, which will contribute to the total FIR luminosity of the source. The $0.9 < z < 1.2$ bin has both the largest total number of sources and the largest range of X-ray luminosities, compared to the lower two bins. 

Figure \ref{fig:Lone_hist_zbins} shows the distributions of 1$\,\mu$m luminosities for the AGN in each redshift bin. As expected, a slight trend of increasing luminosity with redshift is apparent, with the mean 1$\,\mu$m luminosity increasing by $\sim$0.3 dex from the lowest to highest redshift bin and the low-luminosity tail of the distribution falling in the lowest redshift bin. While there is an increase in the $1\,\mu$m luminosity with redshift, this increase is relatively small and within the dispersion of the individual distributions (mean log $L_{\mathrm{1\ts\mu{\rm m}}}/{\rm (erg~s}^{-1}) = 44.02\pm0.56, 44.27\pm0.35, 44.42\pm0.37$ for each bin with increasing redshift). Due to the consistent luminosity of the host galaxies and that the wide range of SED profiles observed in Figure \ref{fig:tot_SEDs} are present in all three redshift bins, we conclude that the wide range of emission properties observed in the sample are not driven by redshift evolution. Therefore we do not divide the AGN into these redshift bins through the rest of the analysis and instead further analyze the emission properties of all AGN uniformly.

\subsection{SED Evolution with X-ray Luminosity} \label{sec: Lx bins}
To analyze the change of the SED profile with the intrinsic power of the AGN, the sample is separated into three bins of intrinsic 0.5--10 keV X-ray luminosity. Figure \ref{fig:SEDs_Lxbins} shows the SEDs separated into these three bins, defined as: $43 <$ log $L_{\rm X} < 43.75$ (602 sources), $43.75 <$ log $L_{\rm X} < 44.5$ (524 sources), and $44.5 <$ log $L_{\rm X}$ (120 sources) where the X-ray luminosity is in units of erg s$^{-1}$. A clear evolution is visible in the median SED across the $L_{\rm X}$ bins with the lowest bin showing the weakest UV and MIR emission relative to the 1\,$\mu$m luminosity. The emission in these two regions increases as the intrinsic X-ray luminosity grows, with the largest $L_{\rm X}$ bin showing an approximately flat median SED in log-log space. As was done in Figure \ref{fig:SEDs_zbins}, the normalization in each bin shown in Figure \ref{fig:SEDs_Lxbins} is scaled by the 1\,$\mu$m normalization in the central bin, showing that the luminosity at all wavelengths also increases with increasing $L_{\rm{X}}$.

While there are clear differences in the median SED, the total range of SED shapes is much more consistent across each $L_{\rm X}$ bin, with each bin showing a dispersion of more than 2 orders of magnitude in both the relative UV and MIR luminosity. This wide range in the slope of the UV and MIR continua shows that all three bins contain both obscured and unobscured AGN. The most significant difference between the bins is in the number of FIR detections and that the lowest $L_{\rm X}$ bin contains few sources with a strong peak in the UV (i.e., 0.1$\ts\mu$m) that can be seen in the other two bins. This shows some correlation in the observed UV luminosity or distribution of obscuring dust, with the intrinsic X-ray luminosity. This is examined in further detail in \S \ref{sec: AGN Emission}. Even with these few differences, it is clear that each bin in X-ray luminosity captures an extensive range of different AGN emission features.

\subsection{AGN Emission} \label{sec: AGN Emission}
The main wavelength regions of interest when analyzing AGN properties are the UV emission originating through the thermal properties of the accretion disk, the MIR emission from the warm dusty torus, and the FIR emission which may be associated with a more distant, cold dust component around the SMBH or with star formation in the host galaxy. We directly compare each of these emission components to the intrinsic 0.5--10 keV X-ray luminosity to determine how the reprocessed emission correlates with the intrinsic power of the AGN.

 \begin{figure}
    \centering
    \includegraphics[width=0.9\linewidth]{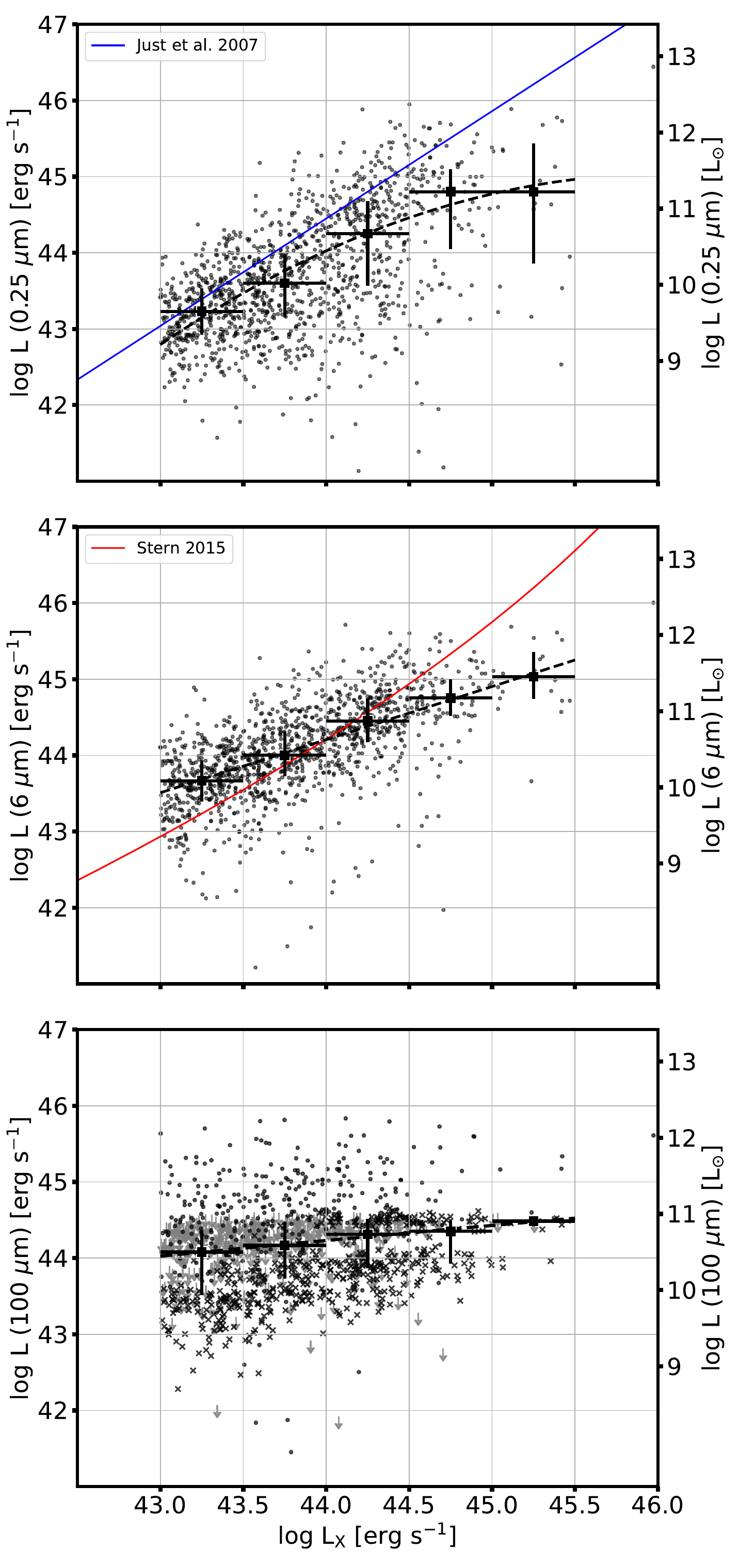}
    \caption{The 0.25$\ts\mu$m (\textit{top}), 6$\ts\mu$m (\textit{middle}), and 100$\ts\mu$m (\textit{bottom}) as a function of the intrinsic 0.5--10 keV X-ray luminosity for all 1246 AGN in the sample. The black squares show the median X-ray luminosity in bins of 0.5 dex along the x-axis. The vertical error bars show the 25--75 percentile spread in each bin and the horizontal error bars show the width of the bin. The black dashed line is a fit to these median points. The blue line shows the relation between the X-ray and UV luminosity from \cite{Just2007}, derived from unobscured AGN. The red line shows the relation between the X-ray luminosity and the $6\,\mu$m luminosity derived from \cite{Stern2015}. The upper limits in the FIR are shown by downward facing arrows and the ``x" show the sources from Stripe~82X that utilize image stacking. The emission at shorter wavelengths (UV-MIR) is AGN dominated in this sample, though the UV is heavily affected by dust. The weaker correlation at 100$\,\mu$m likely means this emission is not entirely dominated by the AGN across the full luminosity range.}
    \label{fig:L_Lx}
\end{figure}

Figure \ref{fig:L_Lx} shows the observed UV (0.25$\,\mu$m), MIR (6$\,\mu$m), and FIR (100$\,\mu$m) luminosity with respect to the intrinsic X-ray luminosity of each AGN. The most luminous AGN that trace the upper bound of the relation seen in the top panel of Figure \ref{fig:L_Lx} show a log-log slope of 0.8$\pm0.06$ and closely match with the relation presented in \cite{Just2007}, which was derived using luminous, unobscured quasars from SDSS. However, there is significant scatter below this relation, even out to high $L_{\rm X}$, with AGN falling more then an order of magnitude below this upper bound. This high dispersion in the relation is likely due to obscuration around the central engine. While the X-ray luminosity shown in Figure \ref{fig:L_Lx} is the intrinsic luminosity that has been corrected for absorption, the UV is the observed luminosity. Therefore, for a given intrinsic $L_{\mathrm{X}}$, we would expect a wide range of observed UV properties, as these AGN are likely to show a wide range in the level of dust obscuration (further discussed in \S \ref{sec: Obscuration 1} and \S \ref{sec: Obscuration 2}). This relation between the X-ray emission and the UV-optical emission in unobscured AGN has also been analyzed in \cite{Steffen2006} and \cite{Lusso2017}, both of which show a slope similar to the high-luminosity sources in this sample.

The central panel of Figure \ref{fig:L_Lx} shows the relation between the intrinsic X-ray luminosity and the MIR luminosity, which is much less affected by extinction and obscuration than the UV luminosity. Correlations between MIR emission and AGN X-ray luminosity have been previously analyzed in many different studies \citep[e.g.,][]{Lutz2004,Fiore2009,Gandhi2009,Lanzuisi2009,Lusso2011,Asmus2015,Stern2015}. The MIR properties of our sample show a log-log slope of 0.7$\pm0.04$ and agree with the relation derived by \cite{Stern2015}, only deviating slightly for the most X-ray luminous sources. This relation is consistent with the MIR--X-ray luminosity correlation showing less scatter than the UV to X-ray correlation, as the MIR emission is significantly less affected by obscuration than the UV emission.

The relation between the FIR luminosity and the intrinsic X-ray luminosity shown in Figure \ref{fig:L_Lx} is more challenging to analyze due to the large number of FIR upper limits of varying depth in each respective field that are present in our sample. A general flat relation in log-log space is seen with large dispersion of more than two orders of magnitude in each direction around the median. However, this dispersion decreases at high $L_{\rm X}$. There are fewer AGN with low FIR luminosity ($L_{\rm FIR} < 10^{43.5}$ erg s$^{-1}$) at  $L_{\rm X} > 10^{44}$ erg s$^{-1}$, showing a lower bound that raises with increasing X-ray luminosity. This slight trend may imply a connection between the intrinsic X-ray luminosity and cold dust emission, with large X-ray luminosites correlating with a lack of low FIR luminosites. However this exact relation depends on many additional factors, such as the dust mass and covering factor of the cold dust. Further discussion concerning how the cold dust component depends on the AGN activity is discussed further in \S \ref{sec:FIR emission}.

\subsection{Obscuring Column Density} \label{sec: Obscuration 1}

\begin{figure}
    \centering
    \includegraphics[width=\linewidth]{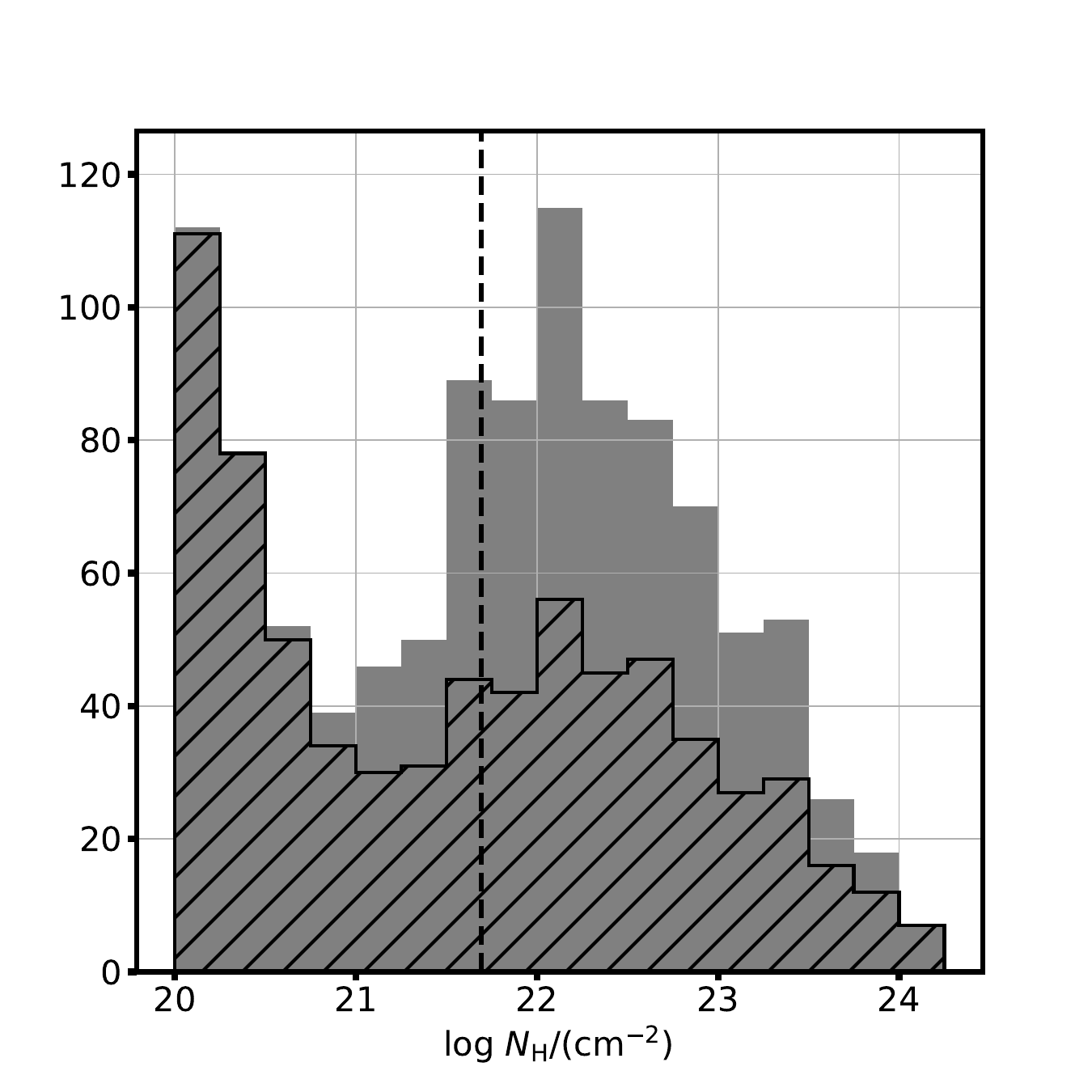}
    \caption{The distribution of obscuring column density, $N_{\mathrm{H}}$, for the measured estimates ($\sim56\%$ of the sample) and the upper limits ($\sim32\%$ of the sample). The total sample is shown in gray while the measured estimates are shown as the hatched histogram. Most show signs of moderate to heavy obscuration.}
    \label{fig:Nh_hist}
\end{figure}

The column density of neutral hydrogen $N_{\rm H}$ is often used as a tracer for the level of obscuration of an AGN and is found through detailed spectral fitting in the X-rays or approximated from the hardness ratio (i.e., the ratio of counts or count rate in the hard and soft X-ray bands). Most detailed analyses of AGN X-ray properties report an estimate of $N_{\rm H}$ for sources with high quality X-ray spectra. However, determining this value accurately is strongly dependent on the quality of the X-ray data, as a large number of X-ray photons are needed to properly determine the spectral shape. This leads to many sources having only upper limits or no reported estimate of $N_{\rm H}$, particularly the fainter sources in the sample. In total about $57\%$ of our sample has a secure measurement of the obscuring column density, while an additional about $32\%$ of the sample has a reported upper or lower limit for $N_{\rm H}$. The distribution of these limits is consistent with the distribution of the confirmed measurements, but lacking a peak at $N_{\rm H} = 10^{20} \rm{cm}^{-2}$, seen in Figure \ref{fig:Nh_hist}. As unobscured AGN are typically assigned to these low column density bins, this peak is likely not physical, but rather represents the number of unobscured AGN in the sample. Each respective X-ray catalog has details about how these measurements and upper limits were made (Stripe~82X: \citealt{Peca2022}; COSMOS: \citealt{Marchesi2016,Marchesi2016b,Lanzuisi2018} GOODS-N: \citealt{Xue2016,Li2020} GOODS-S: \citealt{Luo2017,Li2020}).

Figure \ref{fig:Nh_hist} shows the distribution of the obscuring column density for each of the AGN in our sample with a reliable estimate, or an upper limit. A bimodal distribution is visible with a peak at the $N_{\rm H}$ = 10$^{20}$ cm$^{-2}$. This is often the lowest value found and thus corresponds to an unobscured AGN. The second peak falls just above $N_{\rm H}$ = 10$^{22}$ cm$^{-2}$. This value corresponds to an obscured Compton-thin AGN with moderate amounts of obscuration. The tail of the distribution extends to just above $N_{\rm H}$ = 10$^{24}$ cm$^{-2}$ which is a heavily obscured, Compton-thick AGN. The declining distribution at high $N_{\rm H}$ is likely due to the bias against detecting heavily obscured AGN in X-ray flux-limited surveys, rather than a decline in the actual numbers of heavily obscured AGN \citep{Hickox2018}. Very few Compton-thick sources are detected in the AHA sample, as at this redshift range telescopes that probe higher energy X-rays ($>10\,$keV), such as {\it NuSTAR} or {\it Swift-BAT}, are typically needed to detect low- or moderate-redshift AGN under these extreme levels of obscuration \citep{Koss2016,Lansbury2017}.

\subsection{Bolometric Luminosity Calculation} \label{sec: bolometric luminosity 1}

\begin{figure}
    \centering
    \includegraphics[width=\linewidth]{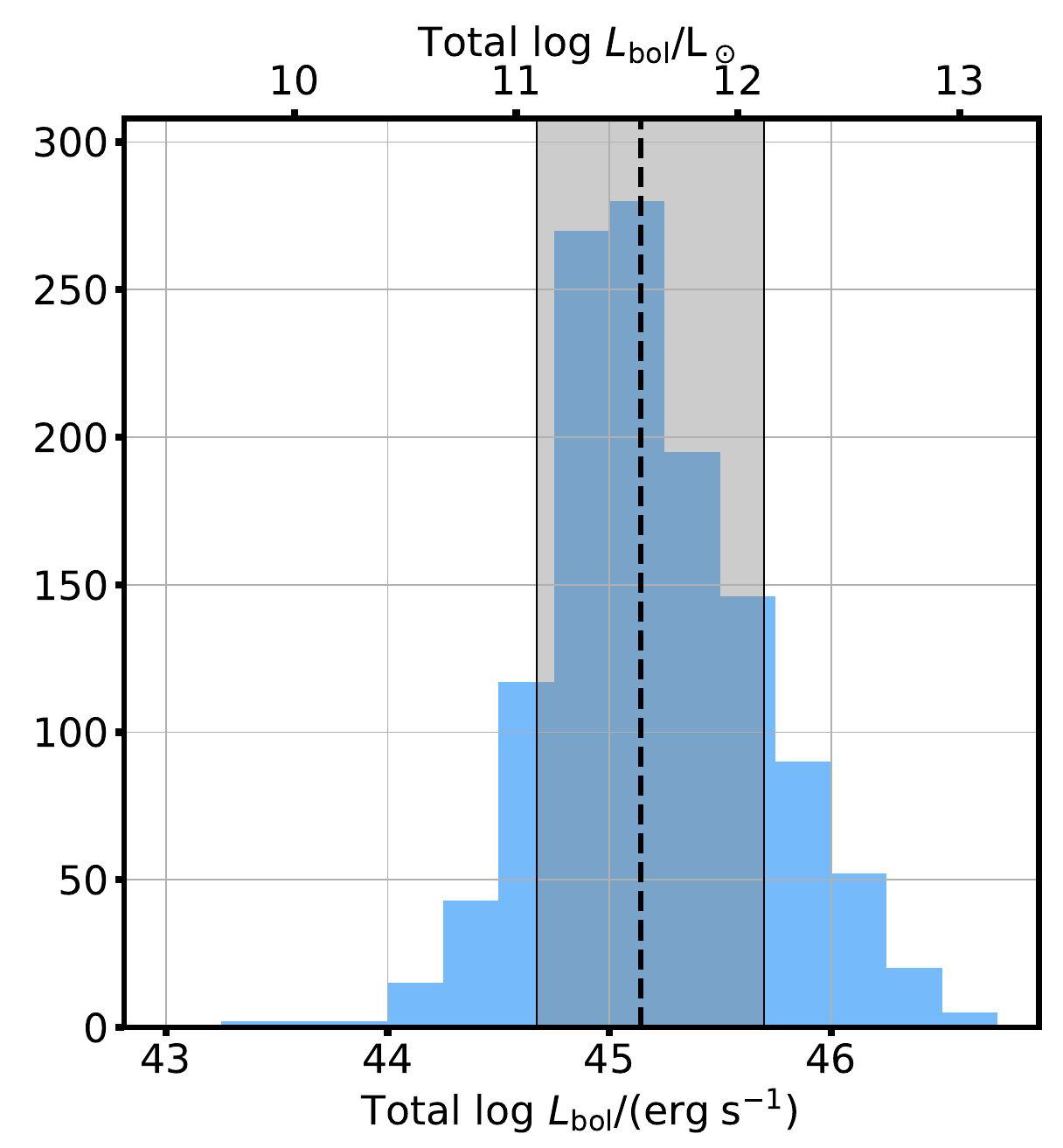 }
    \caption{The distribution of the total bolometric luminosity found by integrating the X-ray to FIR SED of each AGN. The black dashed line shows the mean of log $L_{\rm bol}/(\rm{erg\; s^{-1}}$) = 45.2 distribution and the shaded region shows the 1$\sigma$ spread $\pm0.51$ dex.}
    \label{fig:Lbol_hist}
\end{figure}

The total bolometric luminosity for each source is calculated by integrating the SED from the X-ray to the FIR. The gap in data between the X-ray and UV data is interpolated from the soft X-ray data point and the next shortest wavelength data point of the SED in log space. This region is unobservable due to galactic and atmospheric absorption and the models for the AGN and host galaxy emission are largely unconstrained. However, this linear interpolation provides a reasonable estimate of some contribution from this region without dominating the total bolometric luminosity of the source. For the sources with no FIR detections, the 1$\sigma$ upper limit of the {\it Herschel} SPIRE data or the result from the stacking analysis is used in the calculation of the bolometric luminosity. The FIR can contribute more than 50\% of the total bolometric luminosity for the less luminous sources in the sample, therefore assuming the 1$\sigma$ upper limit in the FIR may overestimate the bolometric luminosity for these sources with no FIR detections. However these low luminosity sources constitute $<10\%$ of our sample without FIR detections and the bolometric luminosity of the more luminous sources, which tend to have the fewest FIR detections, is instead dominated by the UV to MIR emission. Therefore, any upper limits in the FIR will have a much smaller total effect on the bolometric luminosity (approximately 10$\%$ to 25$\%$ of the total luminosity). The distribution of the total bolometric luminosity is shown in Figure \ref{fig:Lbol_hist}.

In order to estimate the bolometric luminosity of the AGN more accurately, a simple elliptical galaxy template from \cite{Assef2010} is subtracted from each source. While this template spans 0.03 -- 30$\ts\mu$m, it only reaches significant levels to contribute to the total luminosity between $\sim$0.3 and $\sim$3$\ts\mu$m.  While AGN are more likely to fall in systems with more star formation and stronger emission over a wider wavelength range than is represented in this template, removing a simple elliptical galaxy allows much of the stellar population from the host galaxy to be removed while ensuring that no UV, MIR, or FIR emission that may be originating from the nuclear region is inaccurately subtracted from the AGN luminosity. Though this may overestimate the AGN bolometric luminosity, particularly the contribution from the FIR, these calculations are not dependent on modeling the star formation histories of these systems.

The elliptical galaxy template is scaled to the $1\ts\mu$m luminosity of each source, up to a maximum luminosity matching the average $1\ts\mu$m luminosity in each redshift bin (see Figure \ref{fig:Lone_hist_zbins}). We do not allow the galaxy template to exceed this value as the most luminous sources likely have contributions from the AGN emission at this wavelength (see $\S$ \ref{sec:5 panel seds} for further discussion of this emission component). For most sources, the elliptical galaxy component is a small fraction of the total bolometric luminosity, however for some of the faint AGN, this stellar component can be the main source of emission. While subtracting the elliptical galaxy component better estimates the total emission of the AGN, star formation within the host galaxy could still contribute significantly to the UV or FIR. Therefore, our integrated estimates can be considered as upper limits to the AGN bolometric luminosity for each system.

A central challenge in determining the AGN bolometric luminosity is to determine how much of the FIR emission is due to star formation and how much is from cold dust around the central nucleus. A more detailed fitting procedure that takes into account full AGN and star formation history models is necessary to properly disentangle the source(s) heating the dust. While such an analysis is beyond the scope of this paper, it is planned future work for a follow-up analysis of this sample. In the remainder of this work, we simply use the upper limit on the AGN bolometric luminosity.

\begin{figure}
    \centering
    \includegraphics[width=\linewidth]{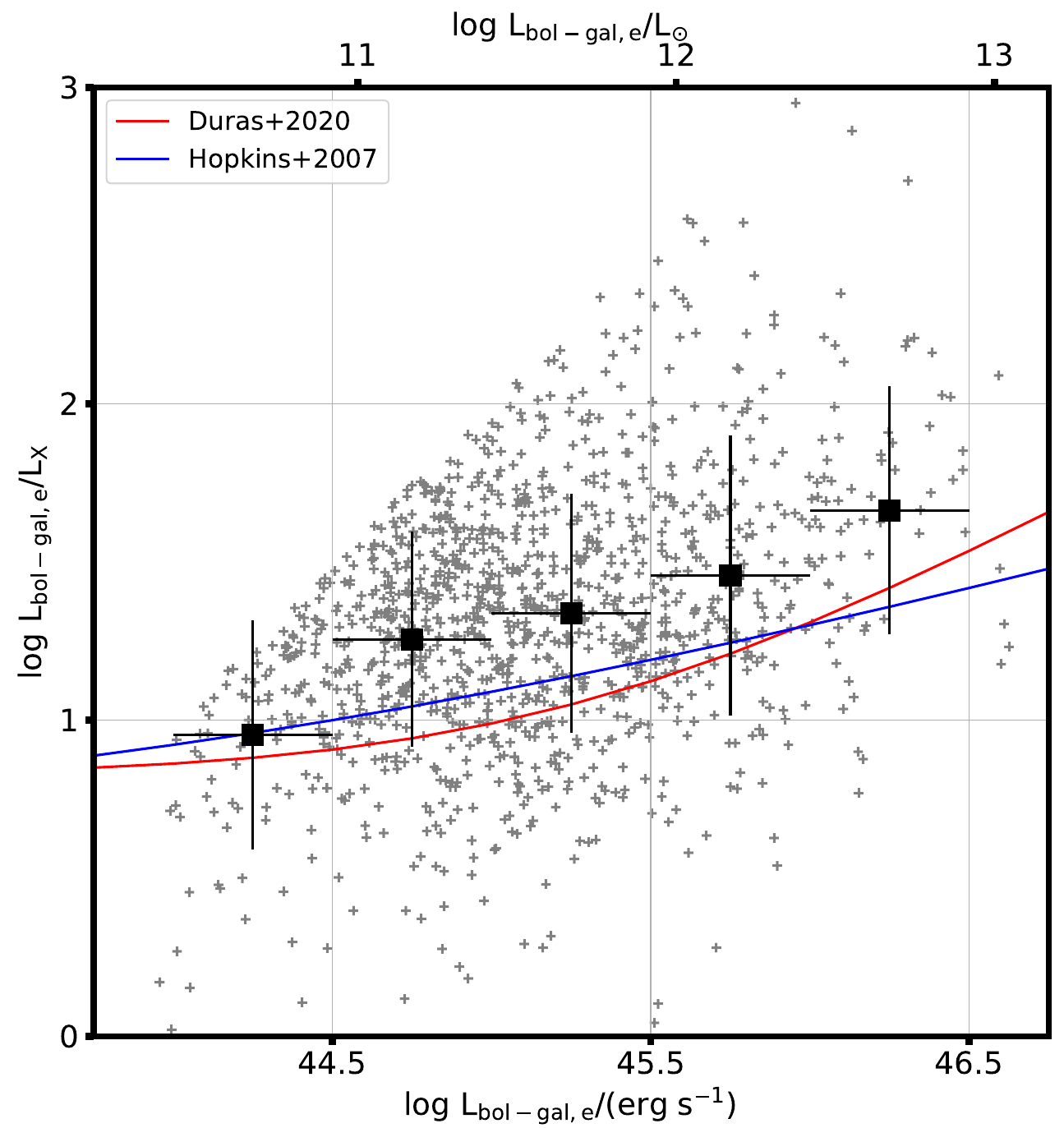}
    \caption{The X-ray bolometric correction, $L_{\mathrm{bol}}$/$L_{\mathrm{X}}$, as a function of the integrated bolometric luminosity. The black squares show the mean in bins of  $L_{\mathrm{bol}}$ with the error bars showing the 1$\sigma$ spread in the y direction and the width of the bin in the x direction. The red and blue lines shows bolometric corrections derived by \cite{Duras2020} and \cite{Hopkins2007}, respectively. The average bolometric luminosity calculated in this work is systematically higher at high bolometric luminosities then the literature values.}
    \label{fig:Lx_Lbol_ratio}
\end{figure}

We compare our bolometric luminosities to the X-ray-to-bolometric corrections for AGN that have been reported in the literature. Figure \ref{fig:Lx_Lbol_ratio} shows the ratio of the calculated AGN bolometric luminosity to the intrinsic X-ray luminosity as a function of the AGN bolometric luminosity. The sharp edge that is present at the upper left of the distribution is due to the lower limit of $L_{\mathrm{X}} > 10^{43}$ erg s$^{-1}$ imposed on the sample. The literature relations from \cite{Duras2020} and \cite{Hopkins2007} are determined by fitting emission models to the SEDs of AGN to calculate the AGN bolometric luminosity. Unlike our calculations these models attribute much of the FIR emission to star formation rather than the AGN. Even with this difference, the literature relations generally agree with our calculated values in showing an increase in the ratio between the AGN bolometric luminosity and the intrinsic X-ray luminosity, with the least luminous sources showing a ratio of $\sim$$10-20$ and the most luminous sources showing a ratio of $\sim$80. While the average distribution in our sample is systematically larger than the bolometric corrections from the literature at high luminosities, there is large scatter in our distribution and the literature relations are consistent within the dispersion. Large scatter in the correction factor was analyzed in \cite{Vasudevan2007} and attributed to different classes of AGN and AGN fitting techniques not taking the fundamental properties of the AGN into consideration, such as the SMBH mass. The offset between the literature values and the bolometric luminosity calculated in this work is also likely driven by the FIR luminosity. \cite{Duras2020} associate all emission beyond $\sim$50$\ts \mu$m with star formation; however, this cold dust component is included in our calculations. 

We can instead fit the SED of the star-forming galaxy M82 (SFR $\sim10\ts M_{\odot}\ts$yr$^{-1}$) to each source, similarly scaling to the 1$\ts\mu$m luminosity in order to approximately match the stellar mass. With this fit, the bolometric correction matches more closely those derived by \cite{Hopkins2007} and \cite{Duras2020}, lowering the median values plotted in Figure \ref{fig:Lx_Lbol_ratio} by an average factor of $\sim$1.5. This effect is most significant for the less luminous sources in the sample, as the more luminous sources have bolometric luminosities driven by the AGN at all wavelengths. Removing the star-forming component by fitting the SED of M82 still leaves the average bolometric correction $\sim$0.15 dex higher on average than the literature values from \cite{Duras2020} for the most luminous AGN.

We further test the contribution of the cold dust component by completely removing the FIR emission for all sources in our sample, by truncating the integration of the SED at 15$\ts\mu$m, removing all FIR emission and any dependency on upper limits or stacking in the L$_{\rm{bol}}$ calculation. When this cold dust component is completely removed from our AGN bolometric luminosity calculations, then our average correction is closer to the \cite{Duras2020} and \cite{Hopkins2007} relations, though still a factor of $\sim$1.25 larger. The remaining offset likely comes from differences in the contributions to the AGN luminosity from the UV.

Without a more advanced fitting technique with a SED fitting routine such as CIGALE \citep{Boquien2019,Yang2022}, it is difficult to determine if the AHA sources with the largest scatter above the literature relations truly have a large contribution in the FIR from the AGN, or if the SED has a significant star-forming component. For now, the present calculations, with only the elliptical galaxy template removed, can be considered to be an upper limit on the AGN bolometric luminosity of the sample that will be used throughout the remainder of this analysis.

\section{Discussion} \label{sec:discussion}
While interesting correlations between X-rays and other AGN components can be seen in Figures \ref{fig:L_Lx} and \ref{fig:Lx_Lbol_ratio}, it can be challenging to disentangle these properties and correlate them with the three dimensional circumnuclear environment around the SMBH.
Here we discuss how the relations between UV and intrinsic X-ray luminosity, and MIR and X-ray luminosity, provide insight into the evolution of an AGN and its circumnuclear dust as its accretion rate changes. A further analysis of observed SED properties is useful to interpret the different properties observed within the AHA AGN sample.

\subsection{Characteristic SED Shapes} \label{sec:5 panel seds}
\begin{figure}
    \centering
    \includegraphics[width=0.845\linewidth]{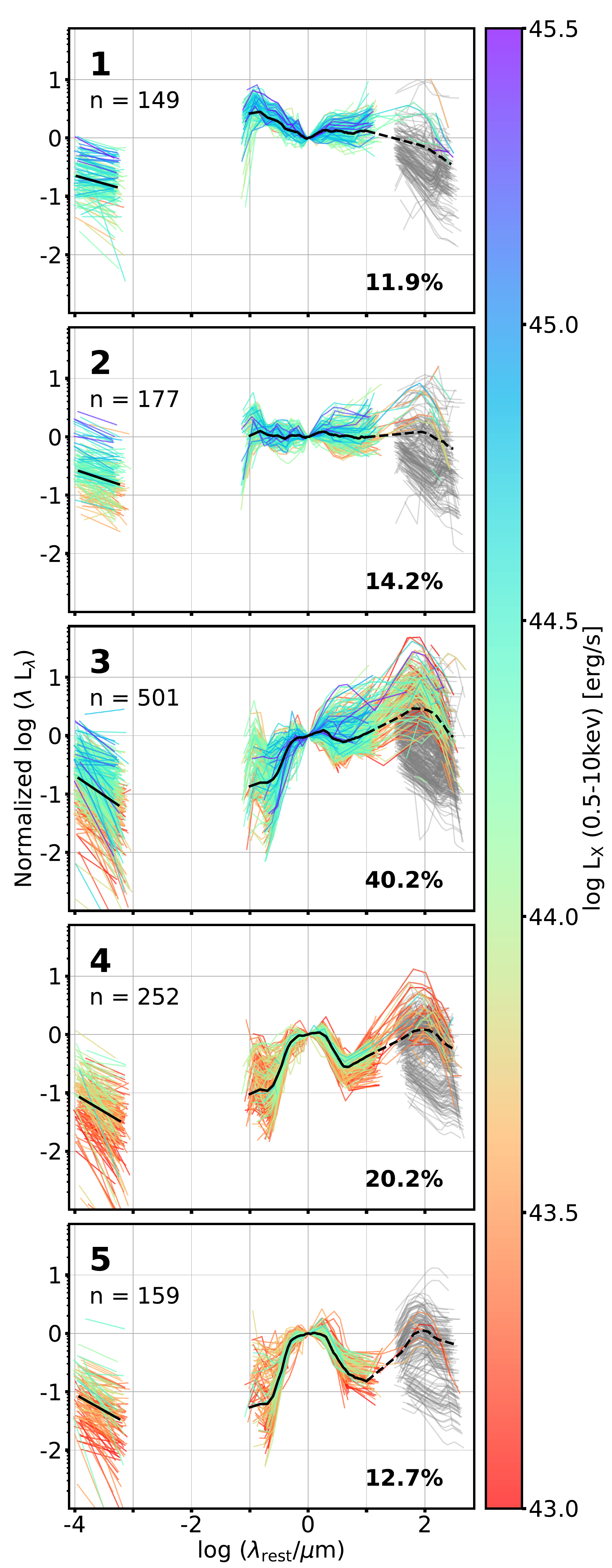}
    \caption{SEDs sorted into five panels based on UV and MIR emission properties (see Table \ref{tab:panel_lims}). Each SED is normalized at rest-frame wavelength 1$\,\mu$m and colored by the intrinsic 0.5--10 keV X-ray luminosity. The numbers indicate the number of AGN in each panel and their percentage of the total sample.
    From top to bottom, the UV and MIR slopes change from falling to rising. Note: 8 sources out of 1246 ($0.6\%$) are not included as they fall outside the continuum slope limits, likely due to noise in the photometric data. Removing these sources has little no effect on the total analysis, as they are such a small fraction of the total sample.}
    \label{fig:5panel_zbins}
\end{figure}

As discussed in \S \ref{sec: Redshift}, a wide range of SED shapes can be seen in all three redshift bins of Figure \ref{fig:SEDs_zbins}, with the UV and MIR emission spanning over two orders of magnitude in relative luminosity. While Figure \ref{fig:SEDs_Lxbins} (SEDs sorted by luminosity) shows trends in SED shape with X-ray luminosity, there is still a wide range of SED emission features present at all luminosities. In order to better characterize the emission properties for such a diverse sample of AGN and more clearly separate individual SED profiles, we now sort the AGN based on the slope of their SEDs in the UV and the MIR, defined as $\alpha$ in Table \ref{tab:panel_lims}. We create five distinct groups designed to capture the full range of SED properties present in the AHA sample, and characterize AGN by the relative strength of accretion disk emission and thermal reradiation from a dusty torus. Table \ref{tab:panel_lims} shows the definitions for each of the five groups and Figure \ref{fig:5panel_zbins} shows the SEDs grouped accordingly. The panels are first defined by the slope of the observed UV continuum, where sources in panel 1 have the strongest UV emission with a steep slope, sources in panel 2 show moderate UV emission with a roughly flat slope, and sources in panels 3, 4, and 5 show weak UV emission. This effectively acts as a proxy for the level of obscuration around the AGN (see further analysis in \S \ref{sec: Obscuration 2}), as the UV emission is generated directly from the central accretion disk and is easily obscured by moderate amounts of dust. As panels 3, 4, and 5 all show similar UV emission, they are further defined by their slope of the MIR continuum, which comes from warm dust heated by the the AGN or young stars in the host galaxy.
 
\begin{deluxetable}{lccc}[t]
\tablecaption{Criteria used to separate SEDs into 5 groups in Figure \ref{fig:5panel_zbins}.}
\tablehead{
\colhead{Fig. \ref{fig:5panel_zbins}} & \colhead{(0.15-1.0$\,\mu$m)} & \colhead{(1.0-6.0$\,\mu$m)} & \colhead{(6.0-10$\,\mu$m)} \\
\colhead{(1)} & \colhead{(2)} & \colhead{(3)} & \colhead{(4)}
} 
\startdata
Panel 1 & $\alpha$ $< -0.3$ & $-0.4 < $ $\alpha$ & - \\
Panel 2 & $-0.3 <$ $\alpha$ $< 0.2$ & $-0.4 <$ $\alpha$  & - \\
Panel 3 & $0.2 <$ $\alpha$ & $-0.4 <$ $\alpha$ & - \\
Panel 4 & $0.2 <$ $\alpha$ & $\alpha$ $< -0.4$ & $0.0 <$ $\alpha$ \\
Panel 5 & $0.2 <$ $\alpha$ & $\alpha$ $< -0.4$ & $\alpha$ $< 0.0$ \\
\enddata
\tablecomments{(1) The row number in Figure \ref{fig:5panel_zbins}. 
(2) UV slope and (3)--(4) MIR slopes used to sort the SEDs. The spectral index, $\alpha$, is defined via $F \propto \nu^{\alpha}$.}
\label{tab:panel_lims}
\end{deluxetable}

Sources in panel 1 of Figure \ref{fig:5panel_zbins} show an SED shape with the ``classical'' or ``quasar-like'' unobscured AGN features, such as a big blue bump (BBB) in the UV and strong dusty torus emission. These AGN also tend to have the highest intrinsic X-ray luminosity and fewest detections in the FIR. Sources in panel 2 show a relatively flat SED shape with slightly lower X-ray luminosity and more FIR detections compared to panel 1. Sources in panels 3, 4, and 5 consist primarily of obscured AGN based on the lack of strong UV emission relative to the emission at 1$\,\mu$m. Sources in panel 3 show MIR emission that is increasing into the FIR, while sources in panel 4 show MIR emission that is decreasing beyond 1$\,\mu$m, before increasing again at $\sim$6$\,\mu$m going into the FIR. Sources in panel 5 show MIR emission that is continuing to decrease from $\sim$6$\,\mu$m to 10$\,\mu$m, before some increase again into the FIR while others potentially continue to decrease or remain constant as they have no FIR detection. The majority of AGN have SED shapes matching those of obscured AGN with high FIR emission (panels 3 and 4), while the most luminous unobscured AGN with quasar-like SEDs (panel 1) constitute the smallest fraction of sources. This is consistent with the previous work that has shown the majority of SMBH growth takes place in obscured AGN \citep[e.g.,][]{Treister2004,Treister2009,Hickox2018,Ananna2019}, and reminds us that optically-selected AGN are far from representative of the full population.

\begin{figure}
    \centering
    \includegraphics[width=\linewidth]{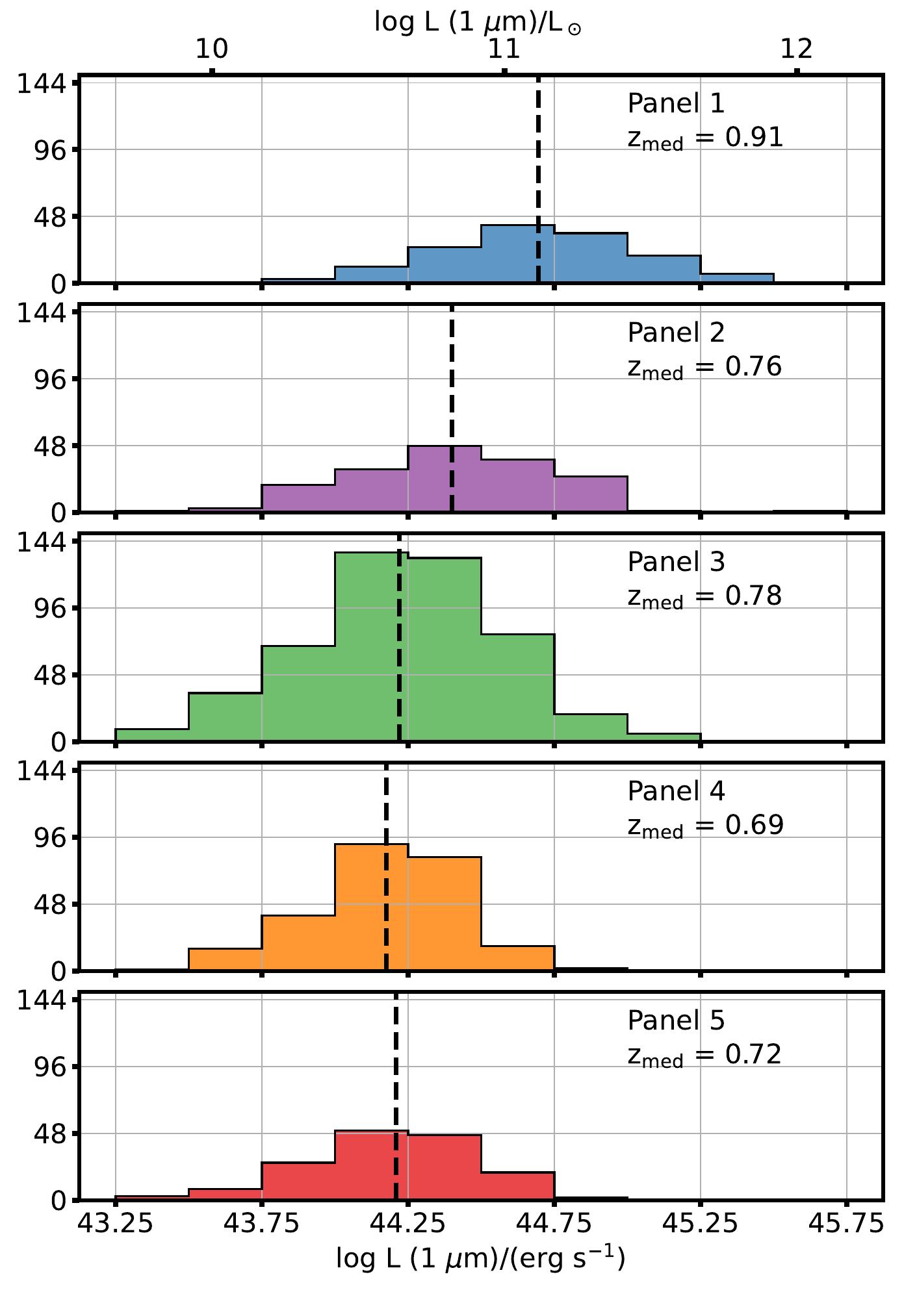}
    \caption{Distributions of 1$\,\mu$m luminosities for the 1246 AGN in our sample, separated according to the 5 SED shapes defined in Table \ref{tab:panel_lims} and shown in Figure \ref{fig:5panel_zbins}. The median redshift of each panel is displayed in each panel and the dashed black line show the median 1$\,\mu$m luminosity of each panel. The sources in panel 1 have 1$\,\mu$m luminosities 3 times higher than the other panels, indicating AGN dominance.}
    \label{fig:Lone_hist_shape}
\end{figure}

 \begin{figure}[h]
    \centering
    \includegraphics[width=\linewidth]{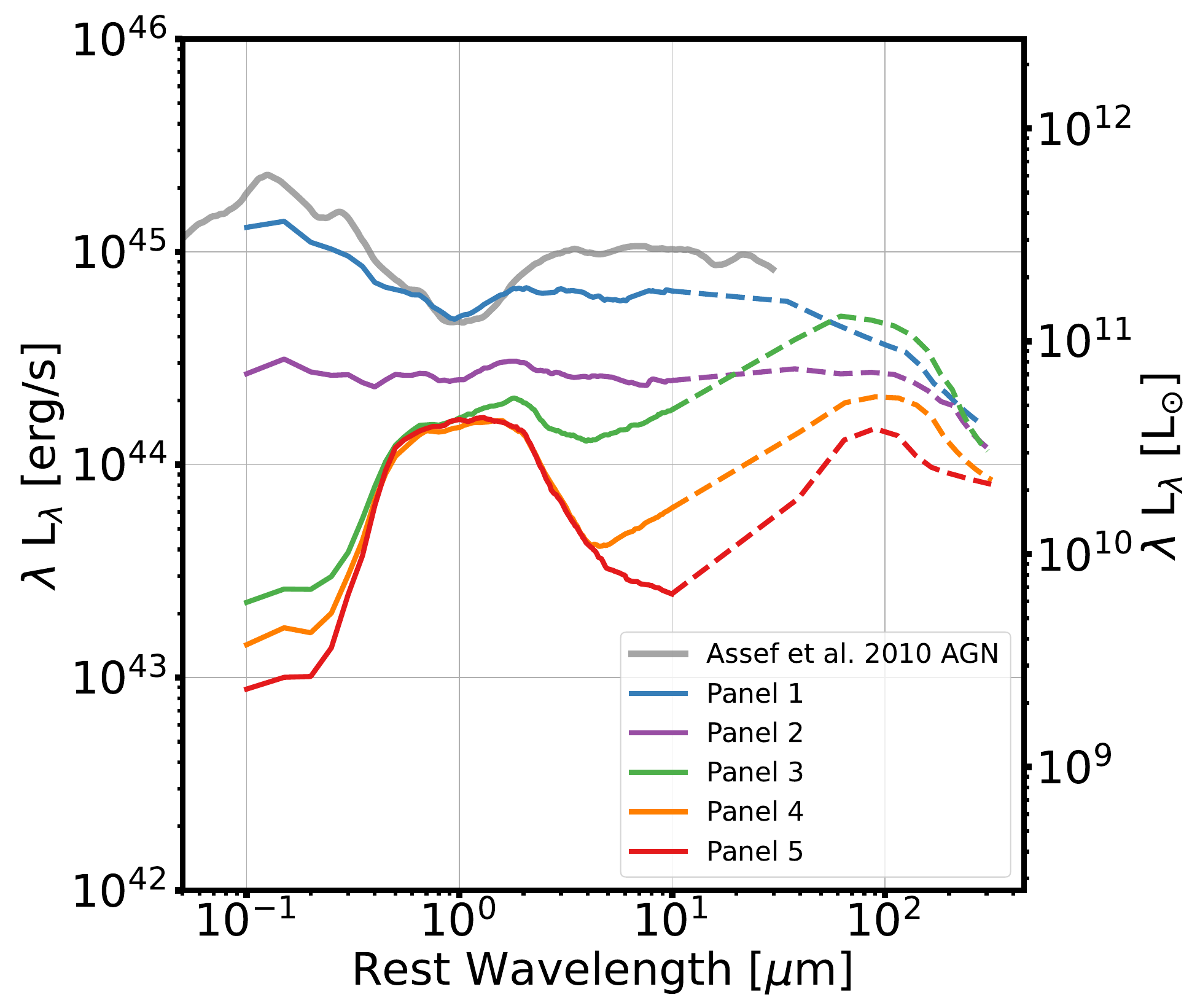}
    \caption{Median SEDs for the 5 panels shown in Figure \ref{fig:5panel_zbins}. The FIR median lines include the 1$\sigma$ upper limits for sources without FIR detections. The gray line shows an unobscured AGN template from \cite{Assef2010}. Sources from panels 3--5 have similar 1$\,\mu$m luminosities, while panel 2 and 1 sources have increasingly larger 1$\,\mu$m luminosities, likely due to contributions from the AGN. Sources from panel 1 are more luminous at all wavelengths from the X-ray to MIR than the sources from the other 4 panels.}
    \label{fig:MedianSEDs}
\end{figure}

While the FIR emission is not utilized when defining these SED panels, interesting insights can be gained based on the number of FIR detections in each panel. The number of FIR detections peaks in panels 3 and 4 and is substantially lower for panels 1, 2 and 5. While panels 1 and 2 show similar emission features with moderate to strong emission in both the UV and MIR, panel 5 shows vastly different properties with the weakest UV and MIR emission. This may imply that the cold dust component of these sources, which generates the FIR luminosity, may be low for two very different stages in the AGN life-cycle, namely, when the luminosity is highest (the quasar phase) and when the luminosity had decreased and the AGN emission no longer dominating the emission of the host galaxy. Where these phases potentially fit in the life-cycle of the AGN is discussed further in \S \ref{sec: bolometric luminosity 2}.

Figure \ref{fig:Lone_hist_shape} shows the distributions of the 1$\,\mu$m luminosity used to normalize the SEDs in Figure \ref{fig:5panel_zbins}. The distributions are roughly constant for the AGN in panels 3--5. However, the sources from panel 2 and panel 1, which resemble increasingly less obscured and more powerful AGN, have an average 1$\,\mu$m luminosity that is 1.5 and 3 times higher than the average for the total distribution respectively. This can also be seen in Figure \ref{fig:MedianSEDs} which shows the un-normalized median SED for each of the five panels.

\begin{figure}
    \centering
    \includegraphics[width=\linewidth]{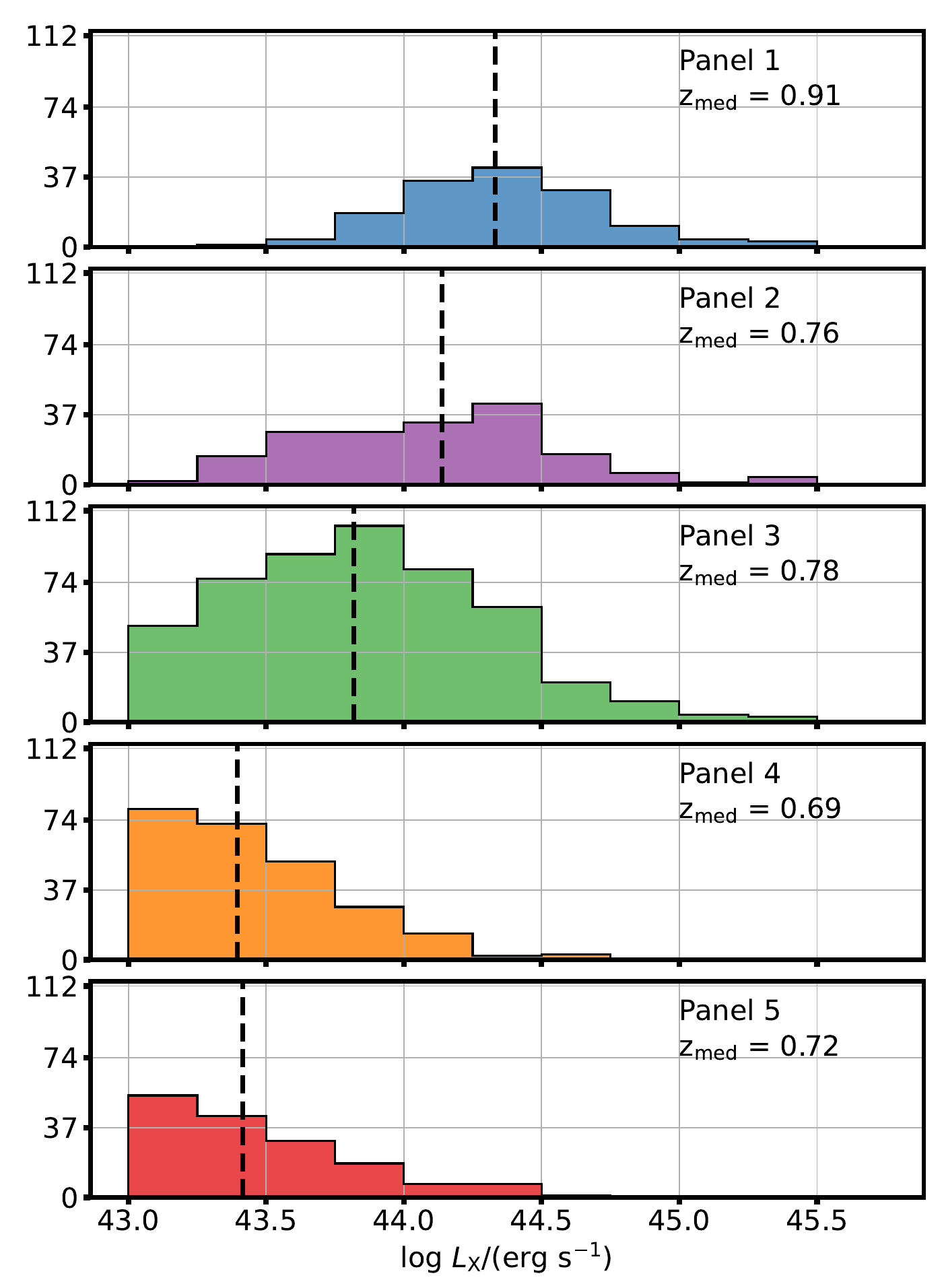}
    \caption{Distributions of intrinsic X-ray luminosity for AGN in each panel from Figure \ref{fig:5panel_zbins}. The dashed black line show the median X-ray luminosity of each panel. A clear trend can be seen with the mean X-ray luminosity decreasing from panel 1 to panels 4 and 5.}
    \label{fig:Lx_box}
\end{figure}

The higher 1$\,\mu$m luminosity for panel 1 and 2 sources is likely caused by strong AGN emission, while for panels 3, 4, and 5, the 1$\,\mu$m luminosity is instead dominated by host galaxy emission. The median 1$\,\mu$m luminosities for the SEDs in panels 3--5 match the average 1$\,\mu$m luminosity for the entire sample shown in Figure \ref{fig:Lone_hist}, showing that the sources that fall in panel 1 and 2 are the outliers on the upper end of this distribution. This steady increase in luminosity from panels 3, 4 and 5, to panel 2, then to panel 1, shows the increase of the AGN emission contribution to emission at all wavelengths.

Assuming that no significant AGN emission is contaminating the 1$\,\mu$m luminosity for panels 3, 4 and 5, we can provide an estimate of the host galaxy stellar mass by following the scaling relation utilized in \cite{U2012} for local galaxies. This leads to an approximate average stellar mass of log ($M_{*}/M_\odot) \sim 10.8 \ts \pm 0.35 \ts$ (FWHM). Such an estimation is much more challenging for the sources in panels 1 and 2 with significant AGN contribution at 1$\ts\mu$m and requires a more detailed SED modeling procedure, which is beyond the scope of this work.

We also find that the redshift distribution for panels 2 -- 5 (the median of which is displayed in Figures \ref{fig:Lone_hist_shape} and \ref{fig:Lx_box}) is roughly constant, with only panel 1 sources being found at preferentially higher redshifts. This may contribute to the increased luminosity of panel 1 sources compared to the rest of the sample, seen in Figure \ref{fig:MedianSEDs}.

Variations in the X-ray luminosity can be seen in Figure \ref{fig:Lx_box}, which shows the distributions of the intrinsic X-ray luminosity for each panel. While the $L_{\mathrm{X}}$ distributions in panels 4 and 5 are largely consistent with one another, the median X-ray luminosity begins to increase into panel 3 and continues to increase through panels 2 and 1, with panel 1 sources showing a median $L_{\mathrm{X}}$ nearly an order of magnitude larger than in panels 4 and 5. As the intrinsic X-ray luminosity this large must be generated directly from the central engine, this trend supports the conclusion that there is likely increased AGN activity in the unobscured sources in panels 1 and 2 compared to those in panels 3, 4, and 5.

\cite{Hao2013} performed an analysis of unobscured AGN in the COSMOS field to define a galaxy-AGN mixing sequence based on the slope of the SED at $0.3-1.0\ts\mu$m and $1.0-3.0\ts\mu$m. The mixing diagram was defined based on these slopes, with the unobscured quasars falling in the bottom right corner of the diagram and sources with more reddening and a lower AGN fraction moving to the right of the diagram and up the diagram respectively \citep[see][for details and additional work]{Hao2013,Bongiorno2012}. While we find general agreement with the interpretation of the five SEDs presented in this work and the previously defined mixing diagram, there are differences in the extent of the AGN contribution and the extent of the reddening by dust compared our sample. We find that panel 2 sources may have the AGN contribution at $1\mu$m overestimated in the mixing diagram as well as panel 4 sources showing more obscuration then may be predicted by \cite{Hao2013}. However, there is general agreement between their proposed mixing-sequence and the possible evolutionary sequence discussed later in this work.

\subsection{Accretion Disk \& Dusty Torus Emission} \label{sec:UV and MIR emission}

\begin{figure}
    \centering
    \includegraphics[width=\linewidth]{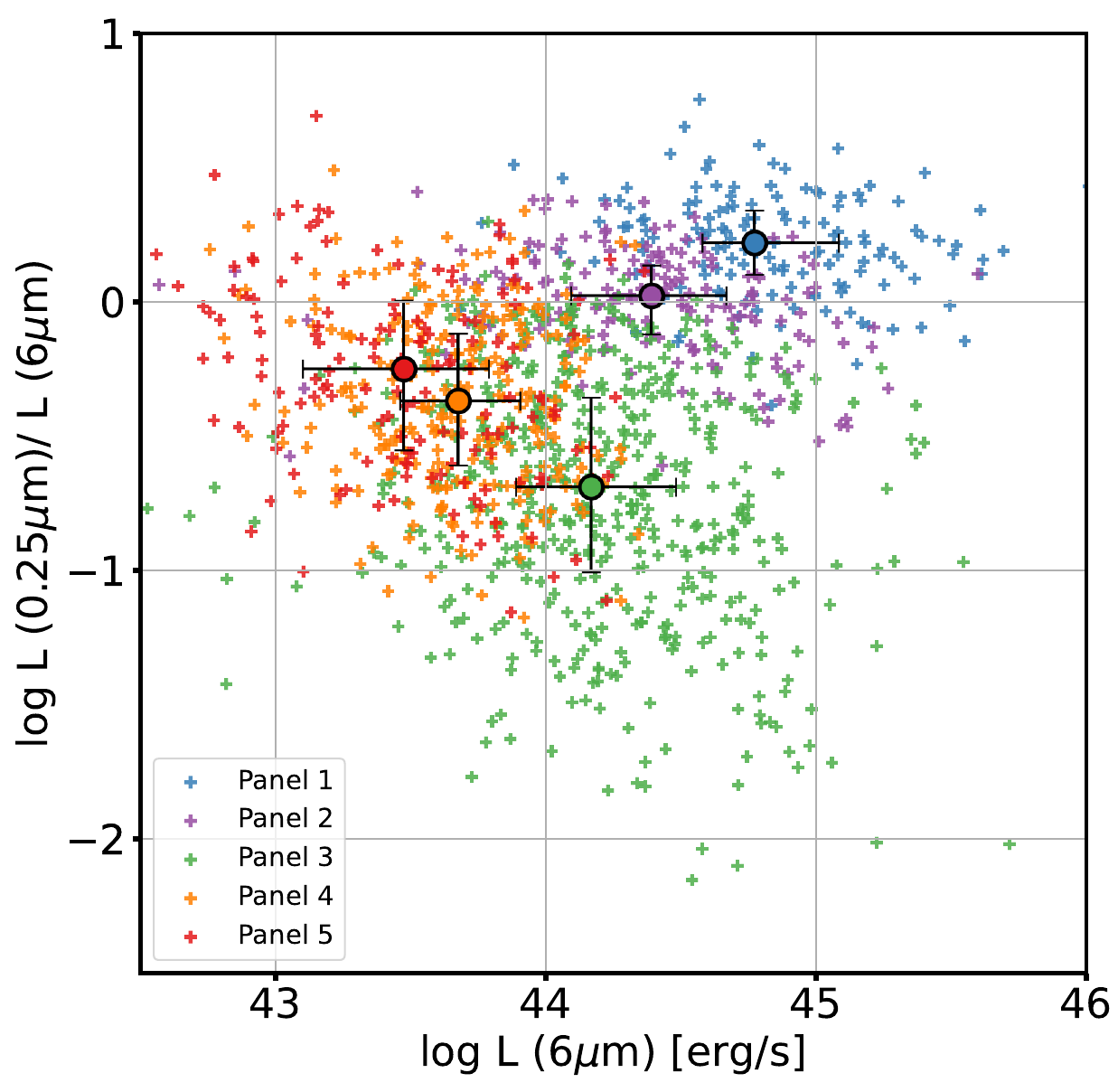}
    \caption{Ratio of UV (0.25$\,\mu$m) luminosity to MIR (6$\,\mu$m) luminosity as a function of MIR (6$\,\mu$m) luminosity. Sources are colored based on which of the 5 panels in Figure \ref{fig:5panel_zbins} they fall into. The median of each of the 5 panels is shown as a large circle with the corresponding color for each panel with the error bars showing the 75th and 25th percentiles of the distribution for each panel. Sources in panel 3 have significantly weaker UV emission for a given MIR emission compared to the rest of the sample, likely driven by obscuration around the SMBH.}
    \label{fig:UV_MIR}
\end{figure}

We are interested in understanding how UV emission from the accretion disk and MIR emission from the dusty torus contribute to the observed AGN SED, and how their relative contributions evolve with the growth of the SBMH. We use the five characteristic SED shapes to place these two emission components in context. Figure \ref{fig:UV_MIR} shows the ratio between UV luminosity at 0.25$\,\mu$m and MIR luminosity at 6$\,\mu$m as a function of the 6$\,\mu$m luminosity, while highlighting which of the five SED shapes each source belongs to. While a general trend of the UV-to-MIR ratio increasing with increasing MIR luminosity is present between the sources in panels 1, 2, 4, and 5, the average location of the sources with SED shapes matching panel 3 in Figure \ref{fig:5panel_zbins} have noticeably lower UV luminosities for a given MIR luminosity. The AGN from panels 1 and 2, the most luminous and unobscured AGN, tend to show the tightest correlation between the MIR and UV luminosity, while sources from panels 3, 4, and 5 also show significantly more scatter. If we are to adopt an evolutionary sequence from panel 4 to panel 1 tracking the increasing intrinsic X-ray luminosity, we can infer that the MIR emission grows more rapidly than the observed UV emission, which is likely hidden behind significant levels of obscuration, leading to the increase in the MIR emission between panels 4 and 3. Once the obscuring dust is removed, potentially blown away by the AGN itself, the observed UV luminosity can increase rapidly, producing the relation seen between the sources in panels 1 and 2.

\begin{figure}
    \centering
    \includegraphics[width=0.84\linewidth]{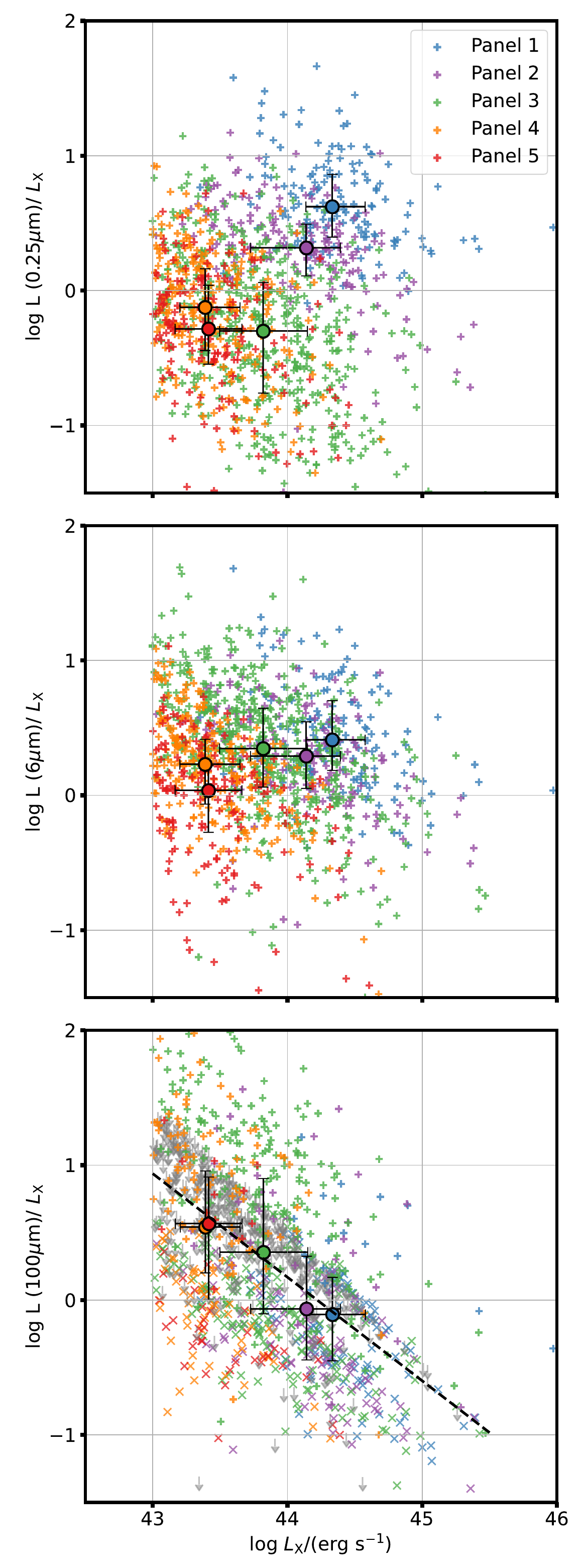}
    \caption{Ratio of UV ({\it top}), MIR ({\it middle}), and FIR ({\it bottom}) luminosity to intrinsic 0.5--10 keV X-ray luminosity as a function of intrinsic 0.5--10 keV X-ray luminosity. Colors correspond to the 5 panels in Figure \ref{fig:5panel_zbins}. The median of each of the 5 panels is shown as a large circle and error bars show the 75th and 25th percentiles of each distribution. The dashed line in the bottom panel is a fit to the data. The upper limits in the FIR are shown by downward facing arrows and the ``x" show the sources from Stripe~82X that utilize image stacking.}
    \label{fig:ratio_Lx}
\end{figure}

The top panel of Figure \ref{fig:ratio_Lx} shows the ratio between UV luminosity at 0.25$\,\mu$m and the intrinsic X-ray luminosity as a function of the intrinsic X-ray luminosity and the middle panel shows the same, but for the relation between the MIR and intrinsic X-ray luminosity. A similar trend to that seen in Figure \ref{fig:UV_MIR} is seen between the UV and X-ray luminosity, with the sources in panels 1 and 2 showing the largest UV luminosity for a given X-ray luminosity and the AGN from panel 3 showing the largest offset from this relation. The large scatter for the more obscured sources in panels 3, 4, and 5 is expected as the UV luminosity will be much more heavily affected by moderate amounts of dust obscuration than the intrinsic X-ray luminosity, which has been corrected for this extinction. While the sources in panel 3 show intrinsic X-ray luminosity similar to those in panel 2, the much weaker UV emission is likely driven by this nuclear obscuration, as was the case in Figure \ref{fig:UV_MIR}. The effect of this nuclear obscuration is further investigated in section \S \ref{sec: Obscuration 2}.

A much more constant relation can be found between the MIR and X-ray emission in Figure \ref{fig:ratio_Lx}, as was also seen in the central panel of Figure \ref{fig:L_Lx}. While large scatter is apparent, the average location of the sources in each SED shape bin is roughly constant. This is to be expected as the MIR dust luminosity of an AGN has been shown to be tightly correlated with the intrinsic X-ray luminosity \citep[e.g.,][]{Lutz2004,Fiore2009,Gandhi2009,Lanzuisi2009,Lusso2011,Asmus2015,Stern2015}. Figure \ref{fig:ratio_Lx} demonstrates that this is true of all AGN within the largely unbiased AHA survey, regardless of the shape of the SED or the slope of the MIR emission. The distribution of all AGN in the central panel of Figure \ref{fig:ratio_Lx} shows significantly less scatter than is present in the UV to X-ray relation and the shape of the SEDs follows a constant relation within the errors of the median. The strength of the MIR correlation with the intrinsic X-ray emission, regardless of the obscuration, shows that much of the IR emission cannot be singularly attributed to star formation, rather much of it is likely powered by the AGN for the most X-ray luminous source (log $L_{\rm X}/({\rm erg \ts s^{-1}}) > 43$). This strong correlation between MIR and intrinsic X-ray luminosity emphasizes the need to determine how much MIR emission is powered by AGN heating and how much by young stars. Failing to account for MIR emission directly associated with the AGN will lead to overestimation of star-formation rate \citep{Kirkpatrick2017,Cooke2020}.

\subsection{Cold Dust Emission} \label{sec:FIR emission}

The FIR emission from cold dust, whether powered by star formation within the host galaxy, by dust heated by the central engine, or by both, can now be analyzed in the context of the total SED shape defined in \S \ref{sec:5 panel seds}. In the bottom panel of Figure \ref{fig:ratio_Lx}, the FIR/X-ray ratios decrease from panels 4 and 5 at low X-ray luminosity to the lower panel 1 ratios at high X-ray luminosity. This strong downward trend is present in both the full sample and when just considering the sources with FIR detections. Figure \ref{fig:ratio_Lx} implies the FIR luminosity must be roughly constant to moderately increasing for increasing X-ray luminosities, as was seen in Figure \ref{fig:L_Lx} and as the SED shapes show (Fig. \ref{fig:5panel_zbins}). \citet{Yang2019} showed a strong correlation for bulge dominated galaxies between the black hole accretion rate and the SFR of X-ray luminous AGN, which traces a connection between the intrinsic X-ray luminosity and FIR luminosity. The much weaker correlation found in our data implies that that AHA sample is not dominated by bulge or spheroidal galaxies, and likely has many more disk galaxies, merging systems, or point sources. A detailed analysis of the host galaxies morphology for each one of these sources is ongoing in the AHA collaboration \citep{Schawinski2014,Powell2017,Ghosh2022,Tian2023}.

Interpreting the exact connection between the AGN activity, which is traced by the intrinsic X-ray luminosity, and the cold dust emission in the FIR is challenging due to the many additional factors that affect this relation. Changes in the dust mass of these systems and the covering factor of dust around the central engine, will directly impact how much of the AGN emission can be reprocessed into the FIR. However, some interesting interpretation can still be made when examining the FIR emission in the context of the total SED shape. The low FIR luminosity and lack of FIR detections for many of the panel 1 and 2 AGN with high X-ray and UV luminosities could be due to the powerful AGN in these sources blowing away cold circumnuclear dust, changing the covering factor, and causing these sources to be less obscured and leading to weaker FIR emission than in the more heavily obscured sources sources found in panels 3, 4, and 5. This is consistent with the finding of \cite{Treister2008,Kirkpatrick2015}, and \cite{Ananna2022,Ananna2022b} and would also imply that a significant component of the FIR luminosity seen in the sources in panels 3, 4, and 5 is directly associated with the AGN, rather than with star formation in the host galaxy. While, the lower relative FIR luminosity seen in panels 1 and 2 could instead be associated with a decrease in the star formation of the host galaxy, perhaps driven by AGN feedback, \cite{Coleman2022} found no direct correlation between the X-ray luminosity and the dust mass of the host galaxy for AGN in Stripe~82X at $z>0.5$.

\begin{figure}
    \centering
    \includegraphics[width=\linewidth]{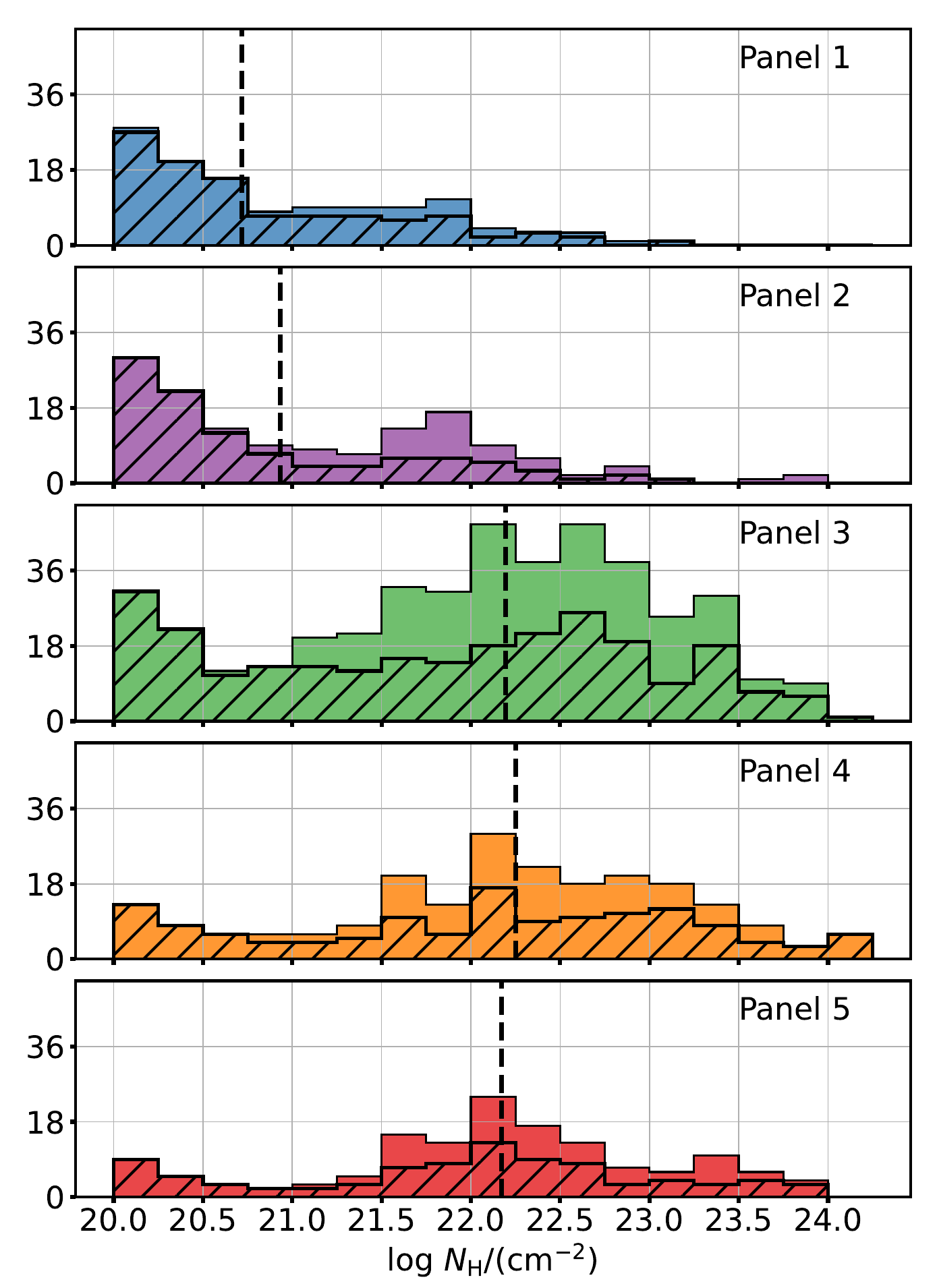}
    \caption{The distributions of the estimated column density, $N_{\mathrm{H}}$, for sources in each panel from Figure \ref{fig:5panel_zbins}. The dashed black line is the median column density in each distribution for each panel. The total sample is shown in each respective color while the measured estimates, without upper limits, are shown as the hatched histograms. AGN from panels 1 and 2 are mostly unobscured while sources in panels 3, 4, and 5 show a wide range of $N_{\mathrm{H}}$ values, from unobscured to Compton-thick.}
    \label{fig:Nh_box}
\end{figure}

\begin{figure*}
    \centering
    \includegraphics[width=\textwidth]{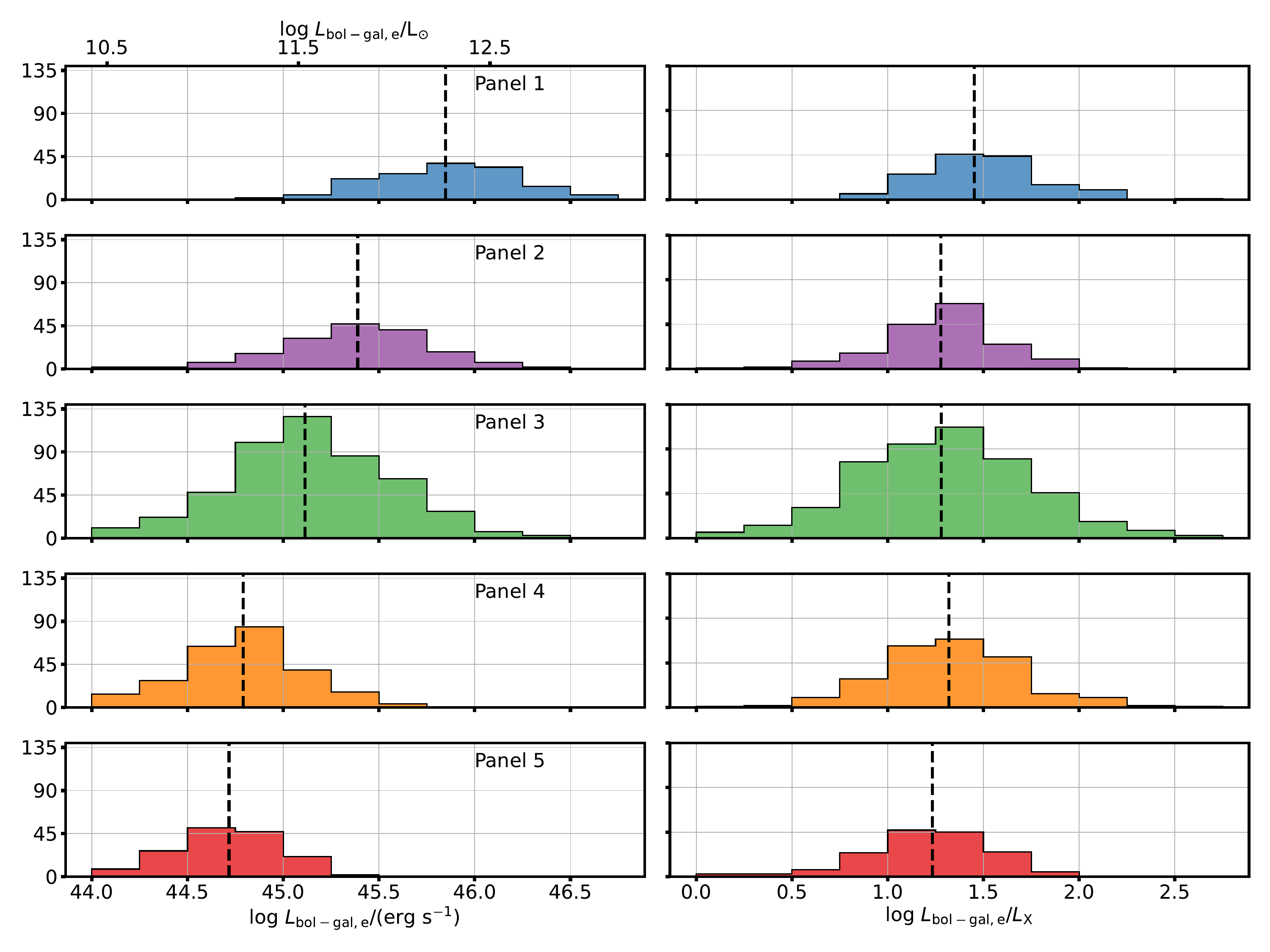}
    \caption{{\it Left:} Distributions of the calculated bolometric luminosity (integrated X-ray to FIR minus the galaxy contribution) for AGN in each panel from Figure \ref{fig:5panel_zbins}. The dashed black line shows the median of each panel. A clear trend can be seen as the median bolometric luminosity decreasing from panel 1 ({\it blue}) to panel 5 ({\it red}). {\it Right:} Ratio between the calculated AGN bolometric luminosity and intrinsic X-ray luminosity as a function of the SED shape defined in Table \ref{tab:panel_lims}. The dashed black line shows the median ratios of each panel. This ratio is remarkably constant, independent of SED shape, indicating that X-ray luminosity is an excellent marker of total AGN power---and that accretion power dominates the energetics of these sources.}
    \label{fig:LxLbol_hists}
\end{figure*}

Ultimately, detailed SED modeling is necessary in order to determine the relative contributions in the FIR from AGN and star formation in the host galaxy. Disentangling these two components is vital to understand the connection between AGN and host galaxy evolution. More sensitive FIR data along with MIR spectroscopy is ultimately necessary to remove the ambiguity caused by the upper limits and to ultimately properly determine the potential connection between the FIR emission and the central engine. Finally, spatially resolving where in the host galaxy the FIR emission is originating from, could unambiguously determine if the cold dust is directly associated with AGN activity.

\subsection{Nuclear Obscuration} \label{sec: Obscuration 2}

The total distribution of the obscuring column density, $N_{\mathrm{H}}$, for all sources that have the high quality X-ray spectra necessary to determine  $N_{\mathrm{H}}$, was shown in \S \ref{sec: Obscuration 1}. We now bin these $N_{\mathrm{H}}$ measurements according to the shape of the SED defined in \S \ref{sec:5 panel seds}.

Figure \ref{fig:Nh_box} clearly shows the connection between estimated column density and SED shape: AGN from panels 1 and 2 have $N_{\mathrm{H}}$ values most consistent with unobscured AGN (the peak of the distribution is $N_{\mathrm{H}}$ $<$ 10$^{22}$ cm$^{-2}$), while AGN from panels 3, 4, and 5 show much higher levels of obscuration, out to the Compton-thick regime ($N_{\mathrm{H}}$ $>$ 10$^{24}$ cm$^{-2}$). Figure \ref{fig:Nh_box} also shows that sources with low UV emission and large MIR emission (panel 3 sources) have the highest fraction of obscured AGN. Though the individual sources with the highest $N_{\mathrm{H}}$ are in AGN showing lower relative MIR emission (panel 4). An analysis of nearby merging galaxies by \cite{Ricci2021} showed that the fraction of heavily obscured AGN peaks when the two galactic nuclei are at a separation of $\sim1$ kpc. Within the context of this work, this may imply that a higher relative fraction of mergers may be present within the sources in panel 3, which show high levels of obscuration, along with large intrinsic X-ray emission (seen in Figure \ref{fig:Lx_box}).

The observed trends discussed here construct an interesting picture of the emission properties of X-ray luminous AGN. The increasing intrinsic X-ray luminosity from panels 4 and 5 to panel 1 (Figure \ref{fig:Lx_box}) and the relative levels of obscuration that peak in panel 3 and sharply decrease in panel 1 and 2 provides additional evidence of a direct connection to the intrinsic power of the AGN (traced by the intrinsic $L_{\rm X}$) and the covering factor of the obscuring dust around the central engine. As the AGN power increases, we find that the level of obscuration decreases. This is once again in agreement with previous work that provides and evolutionary connection between heavily obscured AGN and unobscured quasars \citep[e.g.,][]{Sanders1989,Lawrence1991,DiMatteo2005,Hopkins2006,Oh2015,Ricci2017b,Ananna2022,Ananna2022b}.

\subsection{Bolometric Luminosity with SED Shape} \label{sec: bolometric luminosity 2}

The bolometric luminosity can now be analyzed in the context of SED shape. The left column of Figure \ref{fig:LxLbol_hists} shows the distribution of bolometric luminosities for AGN in each of the 5 panels. The median value increases by more than an order of magnitude between panel 5 and panel 1, with the greatest increase occurring between panel 2 and panel 1. As panel 1 sources are the most luminous in the X-rays and show quasar-like SEDs, this simply indicates that X-ray and bolometric power are correlated. It is also interesting to note that both the median and maximum bolometric luminosities are nearly equivalent for the unobscured sources in panel 2 and the most heavily obscured sources in panel 3. This implies we have recovered well the intrinsic X-ray luminosity, which in turn is correlated with the bolometric luminosity.

\begin{figure}
    \centering
    \includegraphics[width=0.88\linewidth]{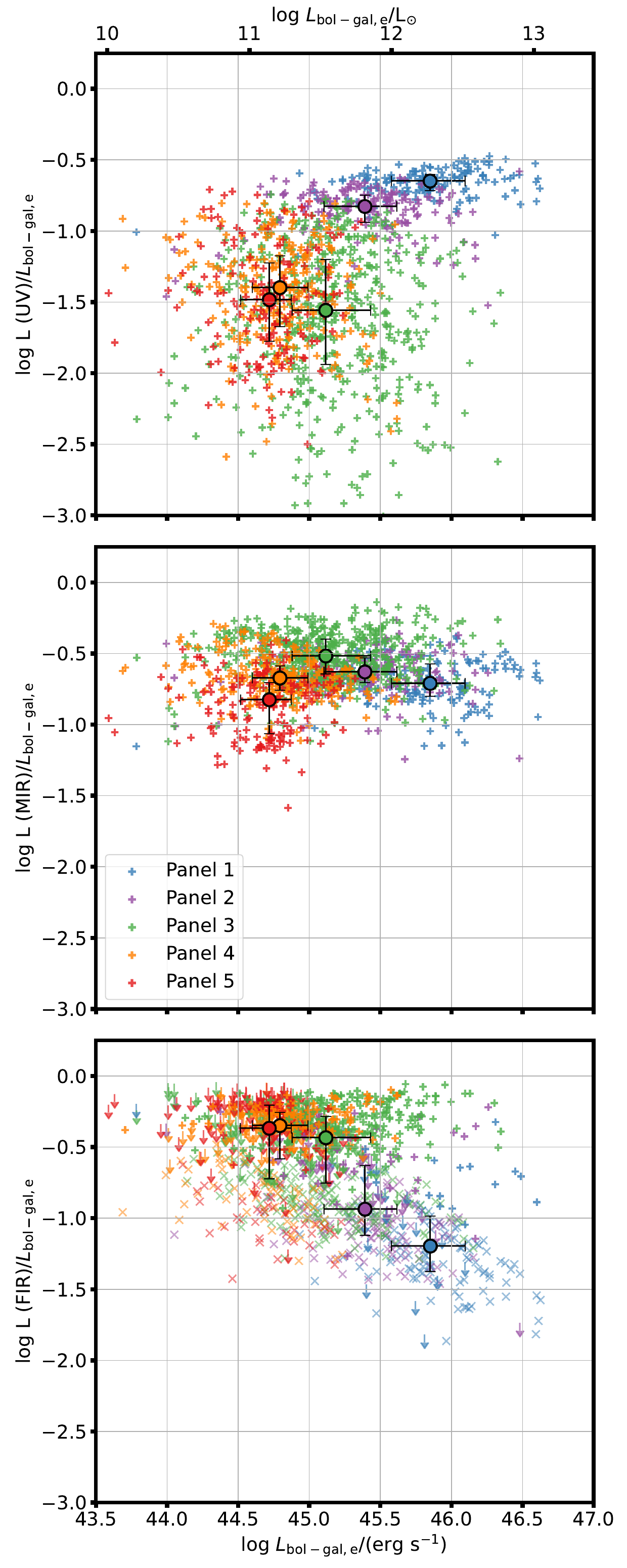}
    \caption{\textit{Top:} Ratio of UV  (0.1-0.36$\,\mu$m) to total bolometric luminosity as a function of bolometric luminosity, for the AHA AGN sample. Sources are colored according to their panel in Figure \ref{fig:5panel_zbins}; medians are shown as large circles; and error bars indicate the 75th and 25th percentiles of the distributions. \textit{Middle:} Same as the top but for MIR luminosity (3-30$\,\mu$m). \textit{Bottom:} Same as the other two panels but for FIR luminosity (30-500$\,\mu$m). The upper limits in the FIR are shown by downward facing arrows and the ``x" show the sources from Stripe~82X that utilize image stacking.}
    \label{fig:Lbol_ratio}
\end{figure}

The right column of Figure \ref{fig:LxLbol_hists} shows the distribution of bolometric corrections from the intrinsic X-ray luminosity ($L_{\rm bol}/L_{\rm X}$) for AGN in each SED shape bin. The values are nearly constant across the 5 SED shapes. This means the growth in bolometric luminosity seen in the left column of Figure \ref{fig:LxLbol_hists} is being driven by the AGN and is directly tied to the growth in the intrinsic X-ray luminosity shown in Figure \ref{fig:Lx_box}. Despite this correspondence, the bolometric corrections are not small---most are in the range 8-100. This tight relation is perhaps surprising, given the wide range of SED shapes across the 5 panels. We can analyze which region of the SED is contributing most to the AGN bolometric luminosity for the AGN in each of the 5 panels to better understand the evolution of the SEDs.
 
Figure \ref{fig:Lbol_ratio} shows the ratios between the total bolometric luminosity of the sources and their UV, MIR, and FIR luminosities, as a function of total bolometric luminosity. It is clear that the bolometric luminosity is mostly dominated by the UV emission from the central accretion disk for the sources from panels 1 and 2. While panel 3 sources have a median bolometric luminosity approaching that of panel 2 AGN, their direct UV contributions to $L_{\mathrm{bol}}$ are very low. This emission is instead reprocessed to MIR and FIR wavelengths, which dominate the contribution to $L_{\mathrm{bol}}$ for panel 3 sources, as well as those in panel 4 and some sources in panel 5.

Figure \ref{fig:Lbol_ratio}, along with the previous several figures, illustrates the similarity in the emission properties of the sources from panels 4 and 5. While these two SED shapes are distinguished by the 10$\ts\mu$m emission, the intrinsic properties (i.e., $L_{\rm X}$, $N_{\rm H}$, and $L_{\rm bol}$) along with their UV (0.25$\ts\mu$m) and MIR (6$\ts\mu$m) emission are remarkably similar. Therefore, at these luminosites, the 10$\ts\mu$m emission is unable to clearly distinguish AGN properties, unlike the other emission features analyzed throughout this work. This is likely due to the silicate absorption features and stronger host galaxy emission at this wavelength.

The transition from UV- to IR-dominated emission at a constant bolometric luminosity---i.e., from panel 3 to panel 2---can be interpreted as further evidence of a possible evolutionary sequence between these 5 panels. As dusty obscured sources become more luminous with further AGN activity (panel 4 to panel 3 transition) the luminous accretion disk may begin to sublimate and blow out some of the obscuring dust, decreasing the overall IR luminosity while increasing the observed UV luminosity (panel 3 to panel 2 transition). Evidence of this removal of dust or receding dusty torus from the most actively accreting SMBHs has been recently shown by \cite{Ananna2022,Ananna2022b} and is in agreement with previous work showing that the obscured fraction of AGN decreases with increasing bolometric luminosity \citep{Lusso2013}. Subsequently, AGN continue to increase in total bolometric luminosity, with the UV component increasing significantly and the cold dust component decreasing (panel 2 to panel 1 transition). AGN with $L_{\mathrm{bol}} >$ 10$^{12.5}$ $L_{\odot}$, which can be seen to the right in the bottom panel of Figure \ref{fig:Lbol_ratio}, may imply further evidence of this obscured to unobscured transition phase. The presence of several sources from panels 1, 2, and 3, at this high bolometric luminosity and high L(FIR)/$L_{\mathrm{bol}}$ fraction, may indicate the start of this transition, just before the cold dust component is removed and the FIR fraction drops, to the locus of the majority of the panel 1 sources. The implication is that much of the FIR emission is directly connected to the central engine, rather than to star formation in the host galaxy.

As previously stated, it is necessary to analyze both the intrinsic AGN and host galaxy properties through the use of detailed SED models that properly account for the AGN emission in the MIR and FIR to further analyze this potential evolutionary sequence. Additionally, the morphology of the host galaxies can shed light on this evolutionary sequence by identifying how it changes over the five characteristic SED shapes. Such an analysis is planned in a future work.

\section{Conclusions} \label{sec:summary}

The combined fields of the AHA wedding cake survey---Stripe~82X, COSMOS, and GOODS-N/S---provide a comprehensive sample of X-ray-selected AGN, which is far less biased against obscured sources than optical or UV surveys. We have utilized the updated COSMOS2020 catalog \citep{Weaver2021} and the most recent Stripe~82X X-ray analysis \citep{Peca2022} to conduct a detailed multiwavelength analysis of X-ray luminous AGN in these three fields out to a redshift of $z_{\rm spec} = 1.2$. We included AGN with intrinsic 0.5--10$\ts$keV X-ray luminosity above $\log L_{\rm X}/({\rm erg\; s^{-1}})>$~43. Targets were carefully selected to ensure that each source has a secure spectroscopic redshift, MIR detection near rest-frame 6$\ts\mu$m, and that they are located in a field with FIR observations from {\it Herschel} SPIRE. This sample selection provides 1246 X-ray luminous AGN for the analysis. When visually inspecting the SEDs for these sources in three redshift bins, we find no major evolution in the multiwavelength properties and therefore do not separate the sources by redshift through the remainder of the analysis. 

The main results of this analysis are:

\begin{enumerate}
    \item The full SED sample shows a wide range of observed rest-frame continuum  strengths (relative to the $1\ts \mu$m continuum), at wavelengths traditionally thought to be powered by the AGN:   3.5 (-3.2, +0.3) dex in the X-ray ($5\ts$keV), 2.6 (-2.0, +0.6) dex in the UV ($0.25\ts \mu$m) and 2.1 (-1.3, +0.8) dex in the MIR ($6 \ts \mu$m). A similar range of relative continuum strengths, 2.8 (-1.2, +1.6) dex is observed in the FIR ($100 \ts \mu$m) when considering the upper limits. 
    
    \item The distribution of rest-frame 1$\,\mu$m luminosity for the full sample is relatively narrow, with a mean  luminosity, log $L_{1\,\mu \mathrm{m}}/({\rm erg~s^{-1}})$ = 44.2 \ts  $\pm 0.46$. Assuming that the 1$\,\mu$m luminosity is dominated by the host galaxy (for all but the most luminous AGN), we conclude that the host galaxies are primarily massive galaxies with mean mass log ($M_{*}/M_\odot) \sim 10.8 \ts \pm 0.35 \ts$ (FWHM). 
    
    \item There is a positive, linear correlation between MIR (6$\,\mu$m) luminosity and intrinsic X-ray luminosity, similar to \cite{Stern2015}. This suggests the MIR emission is largely powered by the AGN. In contrast, only the most luminous AGN in the UV show a positive correlation between the X-ray and UV luminosity. This implies less luminous AGN are far more likely to contain dust that suppresses the UV emission.
    
    \item The calculated bolometric luminosities (X-ray to FIR, with an elliptical galaxy component subtracted) imply bolometric correction factors ($L_{\rm bol}/L_{\rm X}$) increasing from $10-70$ over the range log $L_{\mathrm{bol}}/{\rm (erg~s}^{-1}) = 44.5-46.5$. The trend of this correction is in agreement with \cite{Hopkins2007} and \cite{Duras2020}, but is systematically higher by a factor of $\sim$1.8.
\end{enumerate}

To explore the wide range of SED properties found in this  sample of X-ray luminous AGN, we bin the SEDs based on the relative strengths of the UV (accretion disk) and MIR (dusty torus) emission, which we then relate to the intrinsic source properties, $L_{\rm X}$, $N_{\rm H}$, and $L_{\rm bol}$. We identify 5 SED classes, shown in Figure \ref{fig:5panel_zbins}. SEDs in panel 1 have both strong UV and MIR emission, while panel 5 has both weak UV and MIR emission. Only 11.9\% of the AGN fall in panel 1, with SEDs characteristic of unobscured quasars, 14.2\% show ``flat'' SED shapes (panel 2), 40.2\% show weak UV and strong MIR emission (panel 3), 20.2\% show both weak UV and MIR emission that then shows an increase towards the FIR (panel 4) and 12.7\% show both weak UV emission and MIR emission which then decreases towards the FIR (panel 5). 

The following conclusions refer to relation between these SED shapes and the AGN characteristics:
    
\begin{enumerate}
\setcounter{enumi}{4}
    
    \item The intrinsic X-ray luminosity increases by $1.8\ts$dex from panel 4 and 5 to panel 1, implying that the unobscured, UV-luminous AGN are intrinsically the most powerful AGN. 
    
    \item The ratio $L_{\rm UV}/L_{\rm X}$ increases with intrinsic X-ray luminosity; however, AGN from panel 3 show a significant offset from this relation, with weak UV luminosity for a given X-ray luminosity. Our analysis suggests this is due to increased obscuration in panel 3 AGN. In contrast, the ratio $L_{\rm MIR}/L_{\rm X}$ is constant with increasing X-ray luminosity, which is consistent with the MIR emission being less affected by obscuration. The ratio $L_{\rm FIR}/L_{\rm X}$ decreases with increasing X-ray luminosity, consistent with relatively low to moderate contribution of the $L_{\rm X}$ to the $L_{\rm FIR}$, dependent on the luminosity of the source and plausibly driven by the exact covering factor of cold dust around the central engine.
    
    \item AGN with weak UV emission and strong MIR emission (panel 3) and those with weak UV and weak MIR (panel 4) show the greatest levels and highest fraction of obscuration (mean log $N_{\mathrm{H}} = 22.2$ [cm$^{-2}$]), while sources with strong UV and strong MIR emission show average levels of obscuration, more than an order of magnitude lower (mean log $N_{\mathrm{H}} = 20.7$ [cm$^{-2}$]).
    
    \item AGN bolometric luminosity increases with decreasing panel number, as the obscuration decreases, from low UV and MIR emission (panel 5) to strong UV and MIR emission (panel 1). The ratio between the intrinsic X-ray luminosity and the AGN bolometric luminosity is nearly constant for all SED shapes, with an average value of $20.8\pm1.6$. This is systematically higher then the AGN bolometric correction found by \cite{Duras2020} for all SED shapes.

\end{enumerate}

Evidence for an AGN evolutionary sequence can be seen in the dependence of X-ray luminosity, obscuring column density, and bolometric luminosity on SED shape, as shown by Figures \ref{fig:Lx_box}, \ref{fig:Nh_box}, and \ref{fig:LxLbol_hists}. An X-ray luminous AGN may begin as a low-luminosity, obscured AGN (panel 4 or 5) before increasing in activity and therefore in X-ray luminosity (panel 3). As AGN activity increases further, circumnuclear obscuring dust is removed, allowing the central accretion disk to become unobscured (panels 1 and 2). After the accretion rate and X-ray luminosity decrease, the AGN may return to the panel 5 state once again. This evolutionary sequence is supported by the relative strengths of different AGN components for each SED shape (Figure \ref{fig:Lbol_ratio}). However, this is one interpretation of these SED profiles. It is also likely that not every AGN would undergo this full evolutionary sequence, and may instead live their entire lives as, low luminosity obscured sources (panels 4 or 5). 

In order to fully explore this potential evolutionary sequence, it is vital to better understand how the AGN and host galaxy each contribute to the SED for each of these characteristic SED shapes. This can be explored through SED modeling with programs that properly account for the AGN emission at longer wavelengths, such as CIGALE \citep{Boquien2019,Yang2020,Yang2022}. It is also extremely important to include Compton-thick AGN, which are mostly absent in this sample. Finally, analyzing the morphology of the host galaxy can provide unique insights into the evolutionary state of individual sources. 

The AHA sample of AGN has revealed a far broader range of SED shapes than is seen in optically- or UV-selected samples, because of their bias against obscured sources. Traditional unobscured quasars constitute only about 10\% of the AHA sample---and that is an overestimate given that we are still lacking the most heavily obscured, Compton-thick AGN.  Future AGN models need to explain the wide range of multiwavelength SEDs described here.

\acknowledgments
Acknowledgements: 
We thank the referee for the helpful report that improved the quality of the analysis.
Some of the data presented herein were obtained at the W. M. Keck Observatory, which is operated as a scientific partnership among the California Institute of Technology, the University of California and the National Aeronautics and Space Administration. The Observatory was made possible by the generous financial support of the W. M. Keck Foundation.  
This research is based in part on data collected at Subaru Telescope, which is operated by the National Astronomical Observatory of Japan. We are honored and grateful for the opportunity of observing the Universe from Maunakea, which has cultural, historical and natural significance in Hawai\`{}i. The authors wish to recognize and acknowledge the very significant cultural role and reverence that the summit of Maunakea has always had within the indigenous Hawaiian community.  We are most fortunate to have the opportunity to conduct observations from this mountain. 
Based on observations collected at the European Southern Observatory under
ESO programme ID 179.A-2005 and on data products produced by CALET and
the Cambridge Astronomy Survey Unit on behalf of the UltraVISTA consortium.
C.M.U. acknowledges support from the National Science Foundation under Grant No. AST-1715512, and from NASA through ADAP award 80NSSC18K0418.
We acknowledge support from ANID through Millennium Science Initiative Program - NCN19\_058 (E.T.), CATA-BASAL - ACE210002 (E.T.) and FB210003 (E.T., M. Bo), and FONDECYT Regular - 1190818 (E.T.) and 1211000 (M.Bo.).
TTA acknowledges support from NASA ADAP AWARD 80NSSC23K0486.
M.Ba. acknowledges support from the YCAA Prize Postdoctoral Fellowship.
B.T. acknowledges support from the European Research Council (ERC) under the European Union's Horizon 2020 research and innovation program (grant agreement 950533) and from the Israel Science Foundation (grant 1849/19).
This work has made use of the Rainbow Cosmological Surveys Database, which is operated by the Centro de Astrobiología (CAB/INTA), partnered with the University of California Observatories at Santa Cruz (UCO/Lick,UCSC).
This work is based on observations taken by the CANDELS Multi-Cycle Treasury Program with the NASA/ESA HST, which is operated by the Association of Universities for Research in Astronomy, Inc., under NASA contract NAS5-26555.
This research made use of Astropy, a community developed core Python package for Astronomy \citep{AstropyCollaboration2013,AstropyCollaboration2018}. This research made use of the iPython environment \citep{Perez2007} and the Python packages SciPy \citep{Virtanen2020}, NumPy \citep{harris2020}, and Matplotlib \citep{Hunter2007}. 
\appendix
\section{Sample Data}
\renewcommand\thetable{\thesection.\arabic{table}}    
\setcounter{table}{0}    

We present the data used to generate our SEDs for each of the 1246 AGN in our sample. The tables below list the AHA field each source is located within, the data taken directly from the respective catalogs for each source (i.e., positional coordinates, photometry, X-ray luminosity, redshift, etc.), along with new data calculated in this analysis such as the bolometric luminosity and the characteristic shape of the SED discussed in \S \ref{sec:5 panel seds}.

Specifically, table \ref{tab:appendix_1} lists the physical properties of each source. This includes the AHA field, ID (both from the X-ray and photometry catalog when available), positional coordinates of the X-ray source, intrinsic 0.5--10\,keV X-ray luminosity, column density ($N_{\rm{H}}$), and spectroscopic redshift. The references this information is acquired from is also listed in the table notes and described in more detail in \S \ref{sec:data}. Table \ref{tab:appendix_1} additionally reports the calculated bolometric luminosity with the elliptical galaxy template from \cite{Assef2010} subtracted, as described in \S \ref{sec: bolometric luminosity 1}, and the SED shape classification it falls into based on Table \ref{tab:panel_lims} and Figure \ref{fig:5panel_zbins}.

Table \ref{tab:appendix_2}, \ref{tab:appendix_3}, \ref{tab:appendix_4} list the AHA field, ID, and flux measurements used to create the SEDs for each source analyzed in this analysis in the Stripe~82X, COSMOS, and GOODS-N/S fields respectively. Each field has been separated into different tables as each field has a different set of observational filters making up the SEDs. The sources listed in these tables are the subsets of the total photometry catalogs used in this analysis. The original, complete catalogs for each field are described in \S \ref{sec:data} with the references additionally listed in the table notes. If there is a negative value in a photometric filter, there was either no data reported in the original catalog, or a data point with a high fractional error was reported and subsequently removed from the sample. When possible these negative values are replaced in the analysis by either the upper limits reported in the original catalogs or the flux values resulting from the FIR image stacking analysis in Stripe~82X. It should also be noted that several of the original X-ray catalogs do not report formal errors on the X-ray fluxes. Reasonable values for such measurements were assumed for the analysis, but are not reported in these tables. See the appropriate references listed in the tables and in \S \ref{sec:data} for more details on these X-ray fluxes and their uncertainties.

\begin{deluxetable}{c c c c c c c c c c c}[h]
\tablecaption{Characteristic information of each source in used in this analysis}
\tablehead{
\colhead{AHA Field} & \colhead{X-ray ID} & \colhead{Photometry ID} & \colhead{RA} & \colhead{Dec} & \colhead{$L_{\mathrm{X}}$} & \colhead{z} & \colhead{$N_{\mathrm{H}}$} & \colhead{$L_{\mathrm{bol}}$} & \colhead{Panel} & \colhead{Ref.} \\
\colhead{} & \colhead{} & \colhead{} & \colhead{} & \colhead{} & \colhead{erg s$^{-1}$} & \colhead{} & \colhead{cm$^{-2}$} & \colhead{erg s$^{-1}$} & \colhead{} & \colhead{} \\
\colhead{(1)} & \colhead{(2)} & \colhead{(3)} & \colhead{(4)} & \colhead{(5)} & \colhead{(6)} & \colhead{(7)} & \colhead{(8)} & \colhead{(9)} & \colhead{(10)} & \colhead{(11)}
}
\startdata
        Stripe~82X & 2407 & 2407 & 00h57m17.93907166s & +00d32m44.37456608s & 44.11 & 0.492 & 22.48 & 45.58 & 1 & [1], [2] \\ 
        COSMOS & lid 1217 & 123228 & 09h59m45.0307s & +01d28m28.6445s & 44.27 & 1.029 & 22.05 & 45.48 & 2 & [3], [4] \\ 
        GOODS-N & 164 & 49309 & 12h36m21.22s & +62d11m08.8s & 43.50 & 1.014 & 23.74 & 45.41 & 4 & [5], [6], [9] \\ 
        GOODS-S & 806 & 2000 & 03h32m44.05056s & -27d54m54.4176s & 43.88 & 0.908 & 22.60 & 43.62 & 3 & [7], [8], [9] \\
\enddata
\tablecomments{(1) The AHA field of the source. (2) The ID from the X-ray catalog. (3) The ID from the photometry catalog. (4) Right ascension of the X-ray source. (5) Declination of the X-ray source. (6) Log of the intrinsic X-ray luminosity. (7) The spectroscopic redshift. (8) Log of the reported neutral hydrogen column density. (9) Log of the galaxy subtracted bolometric luminosity. (10) The panel from Figure \ref{fig:5panel_zbins} the source is sorted into. (11) References: [1] \cite{Ananna2017}, [2] \cite{Peca2022} [3] \cite{Weaver2021},[4] \cite{Marchesi2016,Marchesi2016b},[5] \cite{Xue2016},[6] \cite{Barro2019},[7]\cite{Guo2013},[8] \cite{Luo2017}, [9] \cite{Li2020}. This table is available in its entirety in machine-readable form.}
\label{tab:appendix_1}
\end{deluxetable}

\begin{deluxetable}{c c c c c c c c c c c c}
\tablecaption{The Photometric data used to generate the Stripe~82X SEDs}
\tablehead{
\colhead{AHA Field} & \colhead{ID} & \colhead{F$_{2-10{\rm keV}}$} & \colhead{F$_{0.5-2{\rm keV}}$} & \colhead{FUV} & \colhead{FUV err} & \colhead{NUV} & \colhead{NUV err} & \colhead{u} & ... & \colhead{SPIRE 500} & \colhead{SPIRE 500 err} \\
\colhead{(1)} & \colhead{(2)} & \colhead{(3)} & \colhead{(4)} & \colhead{(5)} & \colhead{(6)} & \colhead{(7)} & \colhead{(8)} & \colhead{(9)} & ... & \colhead{(37)} & \colhead{(38)}}
\startdata
Stripe~82X & 2407 & 0.0039  & 0.00266  & 30.169 & 1.492  & 68.007 & 1.3834 & 95.406 & ... & -99.99   & -99.99 \\
Stripe~82X & 2413 & 0.0038  & 0.0137   & -99.99 & -99.99 & -99.99 & -99.99 & 31.941 & ... & -99.99   & -99.99  \\
Stripe~82X & 2420 & 0.00169 & 0.00346  & -99.99 & -99.99 & -99.99 & -99.99 & 2.7939 & ... & -99.99   & -99.99  \\ 
Stripe~82X & 2435 & 0.00033 & 0.00064  & -99.99 & -99.99 & -99.99 & -99.99 & 35.191 &... & 27294.06 & 11242.11  \\
\enddata
\tablecomments{(1) The AHA field of the source. (2) The ID from the photometry catalog. (3-38) The flux and flux errors from the listed band in $\mu$Jy. Photometry comes from \cite{LaMassa2016,Ananna2017,Peca2022} and is described in more detail in \S\ref{sec:s82x} and within each respective paper. Only a portion of the table is shown here. The entire table is available in its entirety in a machine-readable form.}
\label{tab:appendix_2}
\end{deluxetable}

\begin{deluxetable}{c c c c c c c c c c c c}
\tablecaption{The Photometric data used to generate the COSMOS SEDs}
\tablehead{
\colhead{AHA Field} & \colhead{ID} & \colhead{F$_{2-10{\rm keV}}$} & \colhead{F$_{0.5-2{\rm keV}}$} & \colhead{FUV} & \colhead{FUV err} & \colhead{NUV} & \colhead{NUV err} & \colhead{u} & ... & \colhead{SPIRE 500} & \colhead{SPIRE 500 err} \\
\colhead{(1)} & \colhead{(2)} & \colhead{(3)} & \colhead{(4)} & \colhead{(5)} & \colhead{(6)} & \colhead{(7)} & \colhead{(8)} & \colhead{(9)} & ... & \colhead{(45)} & \colhead{(46)}}
\startdata
    COSMOS & 123228 & 9.92E-04 & 0.0041  & -99.99  & -99.99 & 5.71   & 0.1251 & 7.960 & ... & -99.99  & -99.99 \\
    COSMOS & 128954 & 0.00599  & 0.0158  & 1.438   & 0.078  & 4.72   & 0.0919 & 7.716 & ... & -99.99  & -99.99 \\
    COSMOS & 150058 & 0.0013   & 0.0021  & 1.521   & 0.079  & 13.85  & 0.1903 & 19.13 & ... & -99.99  & -99.99 \\ 
    COSMOS & 150214 & 0.0028   & 0.007   & 3.43    & 0.12  & 19.71   & 0.2183 & 17.87 &... & 7451.75 & 731.85 \\
\enddata
\tablecomments{(1) The AHA field of the source. (2) The ID from the photometry catalog. (3-46) The flux and flux errors from the listed band in $\mu$Jy. Photometry comes from \cite{Marchesi2016,Marchesi2016b,Lanzuisi2018,Laigle2016,Weaver2021} and is described in more detail in \S\ref{sec:cosmos} and within each respective paper. Only a portion of the table is shown here. The entire table is available in its entirety in a machine-readable form.}
\label{tab:appendix_3}
\end{deluxetable}

\begin{deluxetable}{c c c c c c c c c c c c}
\tablecaption{The Photometric data used to generate the GOODS-N/S SEDs}
\tablehead{
\colhead{AHA Field} & \colhead{ID} & \colhead{F$_{2-10{\rm keV}}$} & \colhead{F$_{0.5-2{\rm keV}}$ err} & ... & \colhead{F435W} & \colhead{F435W err} & \colhead{B} & \colhead{B err} & ... & \colhead{SPIRE 500} & \colhead{SPIRE 500 err} \\
\colhead{(1)} & \colhead{(2)} & \colhead{(3)} & \colhead{(4)} & ... & \colhead{(11)} & \colhead{(12)} & \colhead{(13)} & \colhead{(14)} & ... & \colhead{(67)} & \colhead{(68)}}
\startdata
    GOODS-N & 48089 & 3.98E-04 & 0.00193 & ... & 5.14 & 0.074 & 10.33 & 0.063 & ... & 12435.0 & 2909.9 \\
    GOODS-N & 49309 & 1.19E-04 & 2.4816E-05 & ... & 0.284 & 0.037 & 0.345 & 0.046 & ... & 6690.0 & 2230.0 \\
    GOODS-S & 37812 & 4.653E-05 & 6.0661E-05 & ... & 2.0975 & 0.06049 & 2.0976 & 0.0605 & ... & -99.99 & -99.99 \\
    GOODS-S & 30253 & 9.306E-05 & 1.4338E-04 & ... & 0.5521 & 0.04247 & 0.5522 & 0.0425 & ... & 18839 & 3070.8 \\ 
\enddata
\tablecomments{(1) The AHA field of the source. (2) The ID from the photometry catalog. (68) The flux the in listed band in $\mu$Jy. The listed photometric IDs reference those from \cite{Yang2014} and \cite{Skelton2014} for GOODS-N and GOODS-S respectively with the photometric data derived from a combination of these catalogs with \cite{Liu2018,Barro2019} and \cite{Elbaz2011,Oliver2012,Guo2013,Hsu2014}. These catalogs are described in more detail in \S\ref{sec:GOODS} and within each respective paper. Only a portion of the table is shown here. The entire table is available in its entirety in a machine-readable form.}
\label{tab:appendix_4}
\end{deluxetable}

\section{FIR Stacking for Stripe~82X} \label{appendix: stacking}
We employ a image stacking in order to achieve more accurate estimates of the FIR emission for the sources undetected in the {\it Herschel} images. For Stripe~82X AGN, sources were grouped in bins of redshift and the observed WISE W4 (22 $\mu$m) luminosity. There are 22 sources with no W4 detections. These sources were placed into a separate bin (Bin 8 in Table \ref{tab:stack}) so the majority of the sources could still be grouped by their MIR luminosity. The {\it Herschel} SPIRE 250 $\mu$m images were then stacked for each bin at the locations of the optical coordinates for each source. The flux measurements were then made through boot-strapping 10,000 runs per bin. This stacking procedure was then repeated on random positions in order determine the error in the stacked detection threshold. This is a similar process to what was done in \cite{Stanley2017} and \cite{Coleman2022}.

\begin{figure}[b]
    \centering
    \includegraphics[width=0.7\linewidth]{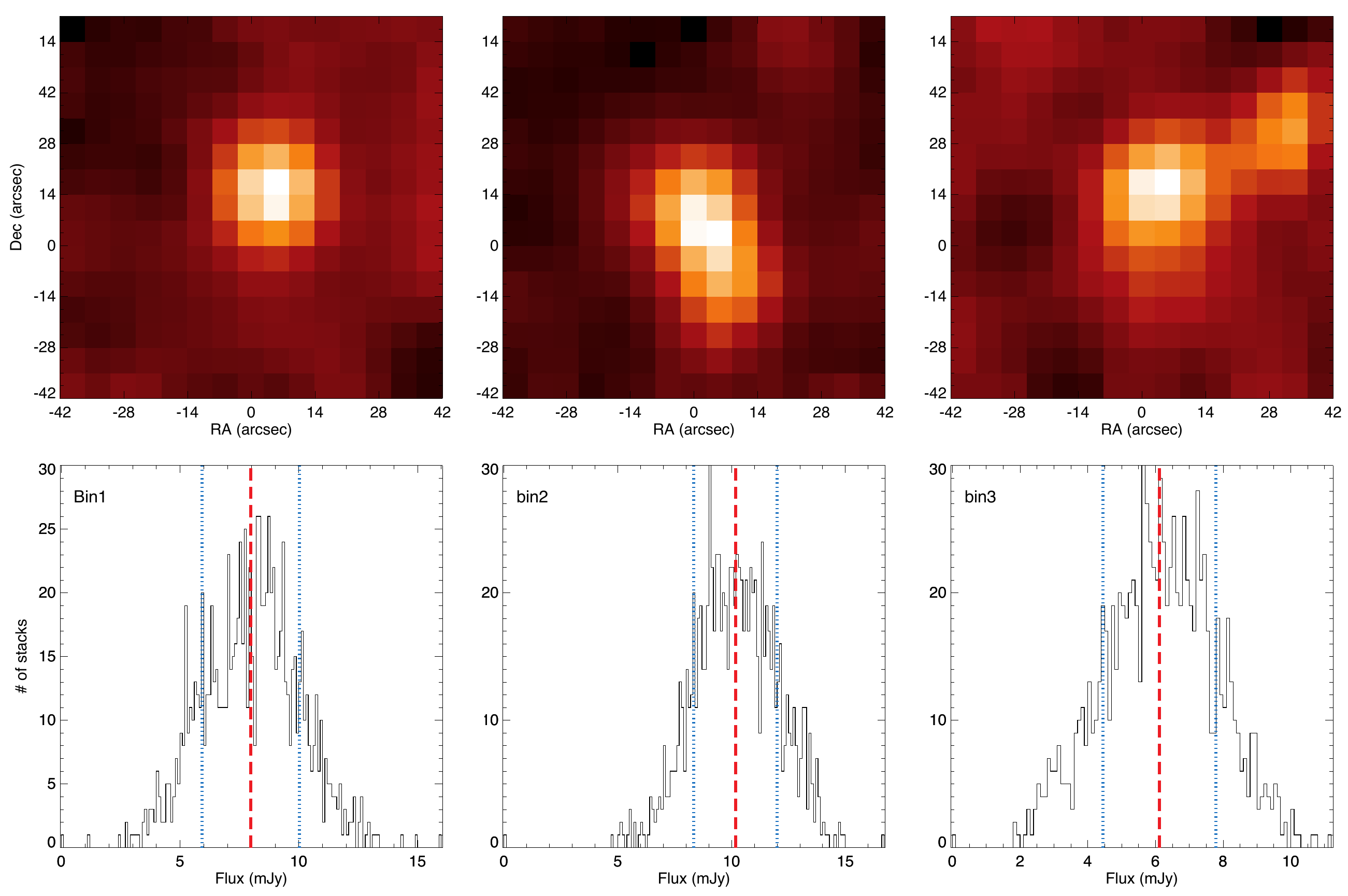}
    \caption{{\it Top:} Stacked images of the first 3 bins from the 250 $\mu$m images. {\it Bottom:} Histograms of the flux measurements made from 10,000 bootstrapped stacks for the corresponding images in the top row.}
    \label{fig:Stacks}
\end{figure}

Table \ref{tab:stack} shows in information for each bin and the results from the stacking analysis. Figure \ref{fig:Stacks} shows three examples of the final stacked images and histograms of the flux measurements resulting from 10,000 bootstrapped stacks. 

\begin{deluxetable}{lcccccc}[b]
\tablecaption{The results from the Stripe~82X 250$\mu$m stacking procedure}
\tablehead{
\colhead{Bin Number} & \colhead{z limits} & \colhead{log L (22$\mu$m) limits} & \colhead{N} & \colhead{Average z} & \colhead{S$_{250\mu {\rm m}}$} & \colhead{dS$_{250\mu {\rm m}}$} \\
\colhead{} & \colhead{} & \colhead{[erg s$^{-1}$]} & \colhead{} & \colhead{} & \colhead{mJy} & \colhead{mJy} \\
\colhead{(1)} & \colhead{(2)} & \colhead{(3)} & \colhead{(4)} & \colhead{(5)} & \colhead{(6)} & \colhead{(7)}
} 
\startdata
Bin 1 & 0.00 -- 0.40 & 0.00 -- 44.0 & 66  & 0.28 & 7.98  & 2.06 \\
Bin 2 & 0.00 -- 0.40 & 44.0 -- 47.0 & 68  & 0.35 & 10.17 & 1.84 \\
Bin 3 & 0.40 -- 0.70 & 0.00 -- 44.5 & 95  & 0.50 & 6.12  & 1.67 \\
Bin 4 & 0.40 -- 0.70 & 44.5 -- 47.0 & 53  & 0.61 & 9.14  & 2.35 \\
Bin 5 & 0.70 -- 1.05 & 0.00 -- 45.0 & 107 & 0.85 & 4.26  & 1.44 \\
Bin 6 & 0.70 -- 1.05 & 45.0 -- 47.0 & 39  & 0.93 & 14.93 & 3.76 \\
Bin 7 & 1.05 -- 1.25 & 0.00 -- 47.0 & 49  & 1.13 & 9.92  & 2.75 \\
Bin 8 & 0.00 -- 1.25 & 0.00 -- 0.00 & 22  & 0.91 & 15.74 & 5.40 \\
\enddata
\tablecomments{(1) The bin number. (2) the redshift range for each bin. (3) The luminosity from the W4 band. (4) Number of sources in each bin. (5) Average redshift of each bin. (6) The resulting 250$\ts\mu$m flux from each bin. (7) The error in the resulting 250$\ts\mu$m found from bootstrapping.}
\label{tab:stack}
\end{deluxetable}

This same process was attempted for the sources in the COSMOS field which lack detections at 250$\ts\mu$m, however, because the depth of the {\it Herschel} images in this field, no signal can be determined that is above the 2$\sigma$ limit of the background noise. A similar problem is found for the sources in GOODS-N/S due to the depth of the {\it Herschel} images and the small number of sources lacking detections at 250 $\mu$m that can be used for stacking in each bin. Therefore no stacking procedure is employed for the sources in COSMOS and GOODS-N/S in the analysis and the upper limits reported in each respective catalog is used instead, as described in \S \ref{sec: constructing seds}.

\bibliography{References.bib}

\end{document}